\documentclass[aps,pra,
nofootinbib,   
reprint,    
groupedaddress]{revtex4-2}

\usepackage{physics}
\usepackage{amsthm}
\usepackage[most]{tcolorbox}
\usepackage{amssymb}

\usepackage{tikz, pgfplots}
\usetikzlibrary{positioning}
\usepackage{hyperref}
\usepackage{orcidlink}

\usepackage{MnSymbol}  

\newtheorem*{theorem*}{theorem}

\makeatletter
\newcommand{\addbar@}[2]{%
  \makebox[0pt][l]{%
    \raisebox{#1}[0pt][0pt]{%
      \kern#2
      $\m@th\mathchar'26$%
    }%
  }%
}
\DeclareRobustCommand{\qbar}{\text{\addbar@{-1.53ex}{0.01em}}q}
\makeatother

\begin{document}


\title{Incompatible observables in classical physics:\\
A closer look at measurement in Hamiltonian mechanics}

%
\author{David Theurel\,\orcidlink{0009-0002-7818-2994}\,}

\thanks{Current address: Helen Wills Neuroscience Institute, University of California, Berkeley, CA 94720. E-mail: \href{mailto:theurel@alum.mit.edu}{\texttt{theurel@alum.mit.edu}}}

\affiliation{Department of Physics, Massachusetts Institute of Technology, Cambridge, MA 02139}


\date{\today}

\begin{abstract}

Quantum theory famously entails the existence of incompatible measurements; pairs of system observables which cannot be simultaneously measured to arbitrary precision. Incompatibility is widely regarded to be a uniquely quantum phenomenon, linked to failure to commute of quantum operators. Even in the face of deep parallels between quantum commutators and classical Poisson brackets, no connection has been established between the Poisson algebra and any intrinsic limitations to classical measurement. Here I examine measurement in classical Hamiltonian physics as a process involving the joint evolution of an object-system and a finite-temperature measuring apparatus. Instead of the ideal measurement capable of extracting information without disturbing the system, I find a Heisenberg-like precision-disturbance relation: Measuring an observable leaves all Poisson-commuting observables undisturbed but inevitably disturbs all non-Poisson-commuting observables. In this classical uncertainty relation the role of h-bar is played by an apparatus-specific quantity, q-bar. While this is not a universal constant, the analysis suggests that q-bar takes a finite positive value for any apparatus that can be built. (Specifically: q-bar vanishes in the model only in the unreachable limit of zero absolute temperature.) I show that a classical version of Ozawa's model of quantum measurement [Ozawa, Phys.~Rev.~Lett.~\textbf{60}, 385 (1988)], originally proposed as a means to violate Heisenberg's relation, does not violate the classical relation. If this result were to generalize to all models of measurement, incompatibility would prove to be a feature not only of quantum, but of classical physics too. Put differently: The approach presented here points the way to studying the (Bayesian) epistemology of classical physics, which was until now assumed to be trivial. It now seems possible that it is non-trivial and bears a resemblance to the quantum formalism. The present findings may be of interest to researchers working on foundations of quantum mechanics, particularly for psi-epistemic interpretations. More practically, there may be applications in the fields of precision measurement, nanoengineering and molecular machines.

\end{abstract}


\maketitle

\section{Introduction}\label{sec_intro}

It is commonly held among the wider physics community that the topic of classical measurement is essentially trivial. I don't mean the modeling in physical detail of any one laboratory setup, which of course can get very complicated, but just the examination of ``measurement'' as a bare-bones physical process, idealized away from as many complications as possible; a theoretical physicist's model of measurement. One way of stating the wide-held intuition is that in classical physics there is in principle no obstruction, on the precision with which one observable can be measured, due to the simultaneous measurement of any other observable. This intuition is in sharp contrast to the situation in quantum physics, where the Heisenberg uncertainty principle (specifically in its ``joint measurement form''~\cite{arthurs1965bstj, arthurs1988quantum, ozawa1991quantum, ishikawa1991uncertainty}) asserts just such a limit. Surely influenced by this attitude, there is a correspondingly sharp contrast between the little attention ever paid to the measurement process in classical physics, and the large attention paid over the decades (deservedly) to that same process in quantum physics. To the best of my knowledge, only a few examples can be attributed to the first category: Heisenberg's own thought experiments in the late 1920's~\cite{heisenberg1949physical} (particularly Heisenberg's microscope); although they served as the motivation for his quantum uncertainty principle, they were essentially classical arguments, augmented only by Einstein's theory of the photon. In 1996 Lamb and Fearn~\cite{lamb1996classical} set up the problem of a classical point particle (the system) in interaction with a second point particle (the ``apparatus'') subject to noise. They stopped short of a thorough analysis; their primary interest being the quantum case. Recently Morgan~\cite{morgan2020algebraic} and Katagiri~\cite{katagiri2020measurement} made use of KvN formalism in independent attempts to use quantum measurement theory to examine measurement in classical mechanics.

The only long-lasting foray into classical measurement seems to be within the body of work surrounding Maxwell's demon; a field known as information thermodynamics~\cite{parrondo2015thermodynamics}. The demon was first conceptualized by Maxwell in 1867~\cite{collier1990two} as a ``very observant and neat-fingered being'' capable of monitoring the molecules of a gas, and, by opening and closing a small door without exerting any work, of sorting the high-energy molecules from the low, thus creating a temperature gradient. This rectifier of fluctuations, if it existed, could then be used to run a perpetual motion machine of the second kind, violating the second law. Writing in 1929 Szil\'ard~\cite{szilard1929entropieverminderung} realized that, if the second law was to hold, somewhere in the demon's monitoring of the molecules (i.e.~in the measurement process) entropy had to be produced. Soon afterwards von Neumann~\cite{von2018mathematical}, in his reading of Szil\'ard, pointed to information acquisition as the key step incurring entropy cost. The latter claim was developed prominently in the 1950's by Brillouin~\cite{brillouin1951maxwell, brillouin1953negentropy} and Gabor~\cite{gabor1961perpetuum}. But in the 1980's Bennett~\cite{bennett1982thermodynamics, bennett1987demons}, building on prior work by Landauer~\cite{landauer1961irreversibility}, argued against Brillouin and Gabor, pointing instead to erasure of the measurement record as the key step incurring entropy cost. This 150-year-long inquiry may be finally nearing a close in recent years, with the answer appearing to be that both sides, Brillouin-Gabor and Bennett, had part of the answer: There is an entropy cost to measurement which prevents the demon from operating as a perpetual motion machine,\footnote{But see~\cite{earman1998exorcist,earman1999exorcist} as to whether this is a conclusion or an assumption of information thermodynamics.} and this cost can be traded between the acquisition and erasure steps~\cite{sagawa2009minimal, sagawa2012thermodynamics, parrondo2015thermodynamics}. 

\begin{figure}
	\centering
	\includegraphics[width=0.5\linewidth]{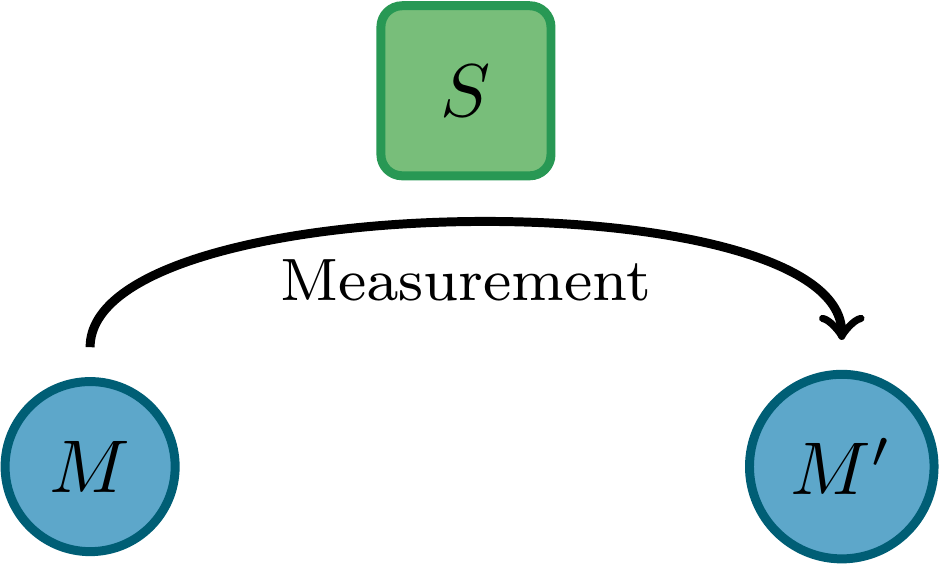} 
	\caption{\textbf{The ideal classical measurement.} A measuring device $M$ extracts information of (is correlated with) a system $S$ without disturbing it. I argue that this is an over-idealization.}
	\label{eq_idealMeasurement}
\end{figure}
With few exceptions, the focus of information thermodynamics has been on determining the costs which measurement, assumed ideal as in Figure~\ref{eq_idealMeasurement}, must incur to safeguard the second law; the focus has not been on the measurement process per se. The question of whether such ideal classical measurements are possible has not, to my knowledge, been seriously addressed. This is exemplified in the popular review by~\textcite{parrondo2015thermodynamics}, which ``[focuses] on measurements where neither the Hamiltonian nor the microstate of the system is affected''. Of the few studies that have explicitly modeled the dynamics of the coupled system and demon, most have used the framework of discrete-state Markov processes~\cite{mandal2012work, barato2013autonomous, mandal2013maxwell, hoppenau2014energetics, strasberg2013thermodynamics, horowitz2014thermodynamics};\footnote{I thank an anonymous referee for pointing out several of these references.} which provides useful effective models, but is too coarse for the ``microscopic'' symplectic structure of phase space and the algebra of observables. \textcite{deffner2013information} considered a Hamiltonian model of a coupled system and memory, but rather than exploring minimally-disruptive measurements, their interest was in leaving the system-memory-bath interaction on indefinitely and obtaining thermodynamic bounds concerning the ensuing steady-state. \textcite{tasaki2013unified} attempted a Hamiltonian analysis of the coupled system and demon, but brushed over the measurement process by repeating the assumption from Figure~\ref{eq_idealMeasurement}, and moved on to a derivation of the Jarzynski and Sagawa-Ueda relations under this assumption. Finally,~\textcite{sagawa2012thermodynamics} and Sagawa and Ueda~\cite{sagawa2008second, sagawa2009minimal} provided a quantum analysis of Maxwell's demon including the measurement process. They claimed that their approach contains the correct classical analysis as a special case, and this claim has been echoed elsewhere~\cite{hasegawa2010generalization}. However, their argument consisted of assuming that classical measurement is non-disturbing, and noting that the quantum formalism includes this as a special case when all quantum operators commute. Hence, their account of classical measurement amounted to repeating the assumption from Figure~\ref{eq_idealMeasurement} without inquiry.

The above review illustrates three points which I would like to contend: (i) Despite the wide-held intuition, measurement in classical physics is far from trivial; (ii) it is a surprisingly underdeveloped subject; and (iii) unacknowledged, it is a subject whose immaturity may have long held back progress in some fields of physics. To address the issue, a reasonable aim would be a theory of measurement in the context of Hamiltonian mechanics, which can be considered the mathematical framework at the foundation of classical physics. The research program I'm suggesting can be summarized as: to systematically bring Bayesian probability to bear on an ontology governed by classical Hamiltonian mechanics, with the full strength, and no more, that is permitted by the geometro-algebraic structure of the ontology. That is; to develop the (Bayesian) epistemology of classical Hamiltonian ontology. The present paper aims to kickstart this program, with no ambition of being the final word. 

I begin by noting that the assumption of perfect information regarding the initial state of the measuring apparatus is unrealistic. In fact it is ruled out as a matter of principle by the third law of thermodynamics; initial uncertainty must be present if for nothing other than for finite-temperature thermal noise. Next I posit a model of the measurement as a physical process. While some minimal assumptions are made concerning the systems that can be used as measuring apparatuses, no restrictions are placed on the system under measurement. This model enjoys substantial generality while at the same time lending itself to Bayesian analysis. It is then shown that, in the process of measurement, the uncertainty in the state of the apparatus propagates into two uncertainties regarding the object-system: one is the imprecision of the measurement; and the other an uncertainty in the magnitude of the back-action caused upon the system; that is, an observer effect. And it is found that these two are bound by a Heisenberg-like precision-disturbance relation. In particular, while I find no obstacle in principle to making a measurement arbitrarily precise, I do find an obstruction to realizing such a measurement without disturbance. Interestingly, in the model the disturbance in question is not arbitrary, but takes the form of time-evolution under the Hamiltonian flow generated by the measured observable; the only thing uncertain is how much ``time'' the system flowed. Thus observables that Poisson-commute with the one measured are spared, while those that do not are disturbed. Later I analyze a second model of measurement, a classical version of Ozawa's quantum model~\cite{ozawa1988measurement}, originally proposed as a means to violate Heisenberg's relation, and find that that model too is bound by the precision-disturbance relation. If this finding were to generalize to all models of measurement, it would mean that in classical physics, like in quantum physics, observables can be simultaneously perfectly-precisely measured if and only if they \mbox{(Poisson-)} commute. 

Next, I derive a novel Liouville-like master equation describing the dynamics of (a rational agent's knowledge of) a system under continuous measurement. This equation, which is analogous to the stochastic master equation appearing in continuous quantum measurement~\cite{jacobs2006straightforward}, is capable of describing more general sequences of measurements including inefficient measurements and simultaneous measurements of multiple observables.

These findings indicate ways in which classical measurement bears a resemblance to the quantum formalism. While I hope the topic will be of interest to several fields of physics, it may be of particular interest to $\psi$-epistemic interpretations of quantum mechanics.

The rest of the paper is organized as follows. I begin in Section~\ref{sec_HamMec} by reminding the reader of the basic concepts and equations of Hamiltonian mechanics. In Section~\ref{sec_measurement} I construct the measurement model and obtain the basic results on which the rest of the paper is based. The precision-disturbance relation is arrived at in Section~\ref{sec_precDistRel}. In Section~\ref{sec_continuousMeasurement} I consider the problems of continuous weak measurement over time, of simultaneous measurements of multiple observables, and of inefficient measurements. In Section~\ref{sec_discussion} I discuss several relevant topics in light of the new results: the similarities, and likely coexistence in the real world, of the classical and quantum uncertainty relations; the epistemic limitations inherent to classical Hamiltonian ontology; a germ of a resource theory for apparatus quality; the subtle interplay between ontology and epistemology in a theory of measurement; and resolutions of some apparent violations of my results. The paper ends by contemplating the road ahead.

\section{Brief recap of Hamiltonian mechanics}\label{sec_HamMec}

Hamiltonian mechanics is a confluence of differential, algebraic and symplectic geometry, Lie algebra and Lie groups. A wonderful resource for the topic is~\cite{arnol1989mathematical}.

Consider a continuous-time dynamical system over a $2n$-dimensional symplectic manifold, called \emph{phase space}. The \emph{observables} of the system (e.g.~position, momentum, angular momentum, etc) are the smooth, single-valued, real-valued functions defined globally over phase space. By convention I take observables to not depend explicitly on time. (With this convention, any explicit time-dependence is regarded as specifying a different observable at each moment in time.) The points in phase space can be expressed in local \emph{canonical coordinates} $(q,p)=(q_1,\dots,q_n,p_1,\dots,p_n)$ (Darboux's theorem). In terms of these coordinates, the state of the system evolves over time according to \emph{Hamilton's equations},
\begin{equation}\label{eq_HamsEqs}
	\dot{q}(t)=\frac{\partial H}{\partial p}(q(t),p(t);t),\ \ \dot{p}(t)=-\frac{\partial H}{\partial q}(q(t),p(t);t),
\end{equation}
where at each moment the system's \emph{Hamiltonian}, $H$, is an observable. Notice that ``Hamiltonian'' and ``$H$'' are indexical terms; they don't specify any concrete function over phase space, but refer to whichever observable happens to serve as the generator of time-evolution (as in~\eqref{eq_HamsEqs}) for a given system at a given time. At each moment Hamilton's equations describe a \emph{flow} $\Phi^H_\tau$ on phase space. Along the integral curves of this flow the value of any observable $A(q,p)$ changes as
\begin{equation}\label{eq_eom}
	\dot A=\{A,H\},
\end{equation}
where $\{A,H\}$ denotes the \emph{Poisson bracket},
\begin{equation}\label{eq_bracket}
	\{A,H\}\triangleq\sum_{j=1}^n\left(\frac{\partial A}{\partial q_j}\frac{\partial H}{\partial p_j}-\frac{\partial A}{\partial p_j}\frac{\partial H}{\partial q_j}\right).
\end{equation}
[Note that~\eqref{eq_eom} follows from~\eqref{eq_HamsEqs} after application of the chain rule to $\frac{d}{dt}A(q,p)$; but also contains~\eqref{eq_HamsEqs} as special cases when $A$ equals one of the canonical coordinates.] Two observables $A,B$ for which $\{A,B\}$ is identically zero are said to \emph{Poisson-commute}, or to be in \emph{involution} with each other. In this case, by~\eqref{eq_eom}, the value of $A$ remains constant along the integral curves of the flow $\Phi^B_t$ (and vice versa). It follows that any observable in involution with the Hamiltonian is a \emph{constant of the motion}. In particular, if $H$ is not explicitly time-dependent then it is itself a constant of the motion (conservation of energy). Including itself, a given observable can be in involution with as few as one and as many as $2n$ independent observables, but only as many as $n$ independent observables can be all in involution with one another. In contrast, if $\{A,B\}=1$ identically then $A,B$ are said to be \emph{conjugate} to each other. In this case $B$ is also said to be ``the'' \emph{generator of translations} in $A$ (and vice versa); because, by~\eqref{eq_eom}, the value of $A$ changes monotonically at unit rate along the integral curves of the flow $\Phi^B_t$. A given observable, $A$, may fail to have a conjugate observable. In this case, in a neighborhood of any regular point of $A$ (i.e.~where $dA\neq0$), it is still possible to speak of a locally-defined conjugate ``quantity'', $B$, which satisfies $\{A,B\}=1$ but fails to satisfy the stringent definition of a bona fide observable. This is illustrated on the 2D phase space by the observable $I=\frac12(q^2+p^2)$ [the Hamiltonian for the simple harmonic oscillator (s.h.o.)]; whose conjugate quantity $\phi=\arg(q+ip)$ (the phase of oscillation of the s.h.o.) either fails to be globally continuous, or else fails to be single-valued, depending on one's choice of definition.

Notice that the components of $(q,p)$ satisfy the \emph{canonical relations}
\begin{equation}\label{eq_ccr}
	\{q_i,q_j\}=\{p_i,p_j\}=0,	\quad	\{q_i,p_j\}=\delta_{ij},
\end{equation}
so each canonical coordinate is in involution with all other coordinates but one, to which it is conjugate. A diffeomorphism of phase space, $(q,p)\mapsto(q',p')$, such that $(q',p')$ again satisfy these canonical relations is said to be a \emph{canonical transformation}. Canonical transformations have Jacobian determinant equal to 1, so they preserve the \emph{Liouville measure} of phase space volume, $d^nqd^np=d^nq'd^np'$. For any flow parameter, $\tau$, the Hamiltonian flow $\Phi^H_\tau$ is an example of an (active) canonical transformation; in particular, Hamiltonian flow preserves the Liouville measure (Liouville's theorem). Changes of coordinates implemented by (passive) canonical transformations are particularly convenient since they preserve the simple form of the Liouville measure, the equations of motion~\eqref{eq_HamsEqs}, and the Poisson bracket~\eqref{eq_bracket}.

\section{A model of measurement in a Hamiltonian world}\label{sec_measurement}

Suppose one wished to measure an observable $A(q,p)$ of the system~\eqref{eq_HamsEqs} at time $t_0$. In the world of Hamiltonian mechanics this can only be done by coupling the system to a measuring apparatus, where the joint system ($=$~object-system $+$ apparatus) is itself a Hamiltonian system, with
\begin{align}
	H_\text{joint}(q,p,x,y;t)&=H(q,p;t)+H_\text{app}(x,y;t)\notag\\
	&\quad+H_\text{int}(q,p,x,y;t).\label{eq_jointHam}
\end{align}
Here $(x,y)$ are canonical coordinates on the $2m$-dimensional phase space of the apparatus; $H_\text{app}$ is the apparatus' Hamiltonian; and $H_\text{int}$ is the interaction between system and apparatus, which I will assume to be switched on only briefly around $t=t_0$. I now stipulate a model for the measurement.

\subsection{System-apparatus coupling}\label{sec_coupling}

Consider the \emph{gauge}, or \emph{pointer display}, of the apparatus; by which I mean the observable of the apparatus which, after interaction with the system, is wanted to reflect the sought-after value of $A$ at time $t_0$. Denote this observable of the apparatus by $P(x,y)$. Suppose $P$ has a conjugate observable, $Q(x,y)$ (so that $\{Q,P\}=1$). For the interaction to imprint the value of $A$ on $P$, the interaction Hamiltonian must involve $A$ and the conjugate quantity to $P$, namely $Q$; because this is the generator of translations in $P$.\footnote{\label{fn_CJL}To be more precise: For any specified pointer $P(x,y)$, by the Carath\'eodory-Jacobi-Lie theorem~\cite{libermann2012symplectic} there exists, in a neighborhood of any regular point of $P$ (i.e.~where $dP\neq0$), a canonical coordinate system for the apparatus in which $P$ is one of the coordinates. By $Q$ I mean the coordinate conjugate to $P$ in this system. The requirement that $Q$ be a bona fide observable amounts to the non-trivial assumption that this coordinate can be extended to a smooth single-valued function globally on phase space. As seen in~\eqref{eq_ccr}, $P$ is in involution with all other coordinates of this system but $Q$. It follows that if, upon expressing $H_\text {int}$ in these coordinates, $Q$ did not appear, then one would have $\{H_\text{int},P\}=0$; and by~\eqref{eq_eom} the interaction would have no immediate effect on the pointer $P$. Since this is the opposite of what is wanted, one sees that $H_\text{int}$ should depend on $Q$.} The simplest interaction of this form is the product
\begin{equation}\label{eq_interactionH}
	H_\text{int}(q,p,x,y;t)=\alpha\delta(t-t_0)A(q,p)Q(x,y),
\end{equation}
where $\alpha$ is a constant of proportionality, and $\delta(t-t_0)$ is the Dirac delta function indicating that the interaction is idealized as taking place instantaneously at $t_0$. Note that this interaction is the classical analogue of that used in the canonical model of quantum measurement~\cite{ozawa1993canonical}, first introduced by~\textcite{von2018mathematical}. I will refer to $Q$ as the apparatus' \emph{probe}.

\subsection{Readying the apparatus}\label{sec_readyApp}

I take a step back to consider how to initialize the apparatus into its ``ready state'' prior to interaction at $t_0$. Being, as one is, in the process of defining what one means by ``measurement'', on pain of circularity one should not appeal to measurement to assess the state of the apparatus, as might be needed to actively manipulate it into a state ready for measurement of the system. This difficulty can be circumvented by letting low-temperature thermalization take care of confining the state of the apparatus to a narrow region of its phase space. The region in question can be specified experimentally by setting up a deep energetic well there---a ``trap''. This trap could be due to a confining gravitational or electrostatic potential; a combination of near-field electric and magnetic fields; a light field; atomic chemical bonds; etc. I write
\begin{equation}\label{eq_trap}
	H_\text{app}(x,y;t)=H_\text{app}^\text{own}(x,y)+\Pi(t)H_\text{trap}(x,y),
\end{equation} 
where $H_\text{app}^\text{own}$ is the apparatus' own, or internal, Hamiltonian, which I take to be time-independent; and $\Pi(t)$ is a rectangular step-function taking only the values $1/0$, describing the on/off switch of the trap. The trap will be switched off for all $t> t_0$; it is only switched on in the time leading up to $t_0$, to help bring the apparatus into its ready state, as will now be described. The trap consists of a deep energetic well which, when switched on ($\Pi(t)=1$), sets the ground state of the apparatus at some point $(x^*,y^*)$ of its phase space. Without loss of generality one may set the coordinates such that $(x^*,y^*)=(0,0)$, and one may assume that the corresponding energy is $H_\text{app}(x^*,y^*)|_\text{trap on}=0$. (Otherwise these conditions can be met by shifted redefinitions of $x,y, H_\text{app}$.) I Taylor-expand $H_\text{app}(x,y)|_\text{trap on}$ around the ground state, obtaining a positive-definite quadratic form:
\begin{align}
	H_\text{app}(x,y)\Big|_\text{trap on}&=H_\text{app}^\text{own}(x,y)+H_\text{trap}(x,y)\notag\\
	&=\frac{1}{2}
	\begin{pmatrix}
		x	&y
	\end{pmatrix}
	\hat M
	\begin{pmatrix}
		x\\
		y
	\end{pmatrix}
	+\mathcal{O}(3),\label{eq_HappWithTrap1}
\end{align}
where $\hat M$ is a symmetric positive-definite $2m$-by-$2m$ matrix of coefficients, and $\mathcal{O}(3)$ denotes all higher-degree terms in the series. As shown by Whittaker~\cite{whittaker1959treatise} (see also theorem by Williamson~\cite{williamson1936algebraic}, explained in~\cite[Appendix~6]{arnol1989mathematical}), there exists a local linear canonical coordinate transformation $(x,y)\mapsto (z,w)$ which reduces~\eqref{eq_HappWithTrap1} to the normal form
\begin{align}
	H_\text{app}(z,w)\Big|_\text{trap on}&=\frac{1}{2}\sum_{i=1}^{m}(b_i^2z_i^2+w_i^2)+\mathcal{O}(3).\label{eq_HappWithTrap2}
\end{align}
Here $b_1\geq b_2\geq\dots\geq b_m>0$ are constants with physical dimensions of angular frequency; they are the natural frequencies of oscillation of the apparatus around its trapped ground state. 

Now to ready the apparatus: While the trap is on, the apparatus is brought into contact with a thermal bath at some temperature $T=1/\beta k_B$, allowed to equilibrate, and then isolated again.\footnote{Instead of removing the bath, one might require just that its coupling to the apparatus be weak enough that it doesn't spoil the measurement record, $P$, on the timescales of interest.} After this one's knowledge about the state of the apparatus is given by the Boltzmann probability distribution
\begin{equation}\label{eq_canonicalDist1}
	\rho(z,w)\,d^mzd^mw\propto e^{-\beta H_\text{app}(z,w)|_\text{trap on}}\,d^mzd^mw.
\end{equation}
Note that in the time between isolation from the bath and measurement at $t_0$ the evolution of the apparatus will preserve this distribution, as opposed to spoiling the preparation, since $H_\text{app}|_\text{trap on}$ is constant under the phase-space flow generated by itself and such flow preserves the Liouville measure $d^mzd^mw$.

At this point I make three requirements that constrain the apparatuses, traps, and temperatures allowed by the model. (i) I require that the trap be harmonic enough, or the temperature be low enough, that in the Boltzmann distribution~\eqref{eq_canonicalDist1} the higher-degree terms in~\eqref{eq_HappWithTrap2} can be neglected. (ii) I require that at least one of the coordinates $w_i$ be in involution with $H_\text{app}^\text{own}$. Let $i=i^*$ be the index of this special coordinate. (If given a choice, one wants the associated frequency $b_{i^*}$ to be as large as possible, for a reason to be seen in Section~\ref{sec_precDistRel}.) The condition means that $w_{i^*}$ will be a constant of the motion of the apparatus when the trap is switched off---a desirable property for the pointer $P$ (introduced in Section~\ref{sec_coupling}); so that the measurement record is stable after the interaction has past. I thus identify the pointer $P\triangleq w_{i^*}$ and the probe $Q\triangleq z_{i^*}$. I denote the corresponding frequency by $\Omega\triangleq b_{i^*}$. Note the physical interpretation of $\Omega$ as the natural frequency of oscillation of the probe around its trapped state. Since $Q,P$ are required to be observables, in making these identifications I am implicitly making the assumption (iii): The pair of conjugate local quantities $(z_{i^*},w_{i^*})$ are globally extendable to smooth single-valued functions on phase space. 

From now on $Q,P$ are the only observables of the apparatus with which I will be concerned. With the above requirements met, one can easily marginalize over all other variables in~\eqref{eq_canonicalDist1} to find the probability distribution over the probe and pointer:
\begin{equation}\label{eq_canonicalDist}
	\rho(Q,P)dQdP=\frac{\beta\Omega}{2\pi}\exp{-\frac{\beta\Omega^2}{2}Q^2-\frac{\beta}{2}P^2}dQ\,dP.
\end{equation}
This is the apparatus ready state. It describes a preparation in which the probe and its conjugate have been set independently to zero, but there remains some uncertainty on their exact values.

\subsection{Integrating Hamilton's equations}\label{sec_integratingHam}

Integrating Hamilton's equations for the joint system, the effect of the interaction~\eqref{eq_interactionH} is to instantaneously change the state of both object-system and apparatus as\footnote{To do this calculation it helps to approximate the $\delta$ by a square impulse of width $\Delta t$ and height $1/\Delta t$. As $\Delta t$ is taken smaller and smaller, the joint Hamiltonian~\eqref{eq_jointHam} becomes dominated by $H_\text{int}$ during the interaction, so that $H$ and $H_\text{app}$ can be neglected during the brief time $\Delta t$. Noting that both $A$ and $Q$ are constant under the flow generated by the interaction Hamiltonian~\eqref{eq_interactionH}, both parts of~\eqref{eq_IntegrateHamsEqs} then follow readily. An analogous quantum calculation can be found, e.g. in~\cite{ozawa1993canonical}.}
\begin{subequations}\label{eq_IntegrateHamsEqs}
\begin{align}
\begin{pmatrix}
	q\\
	p
\end{pmatrix}_{t_0^+}
&=
\Phi^A_{\alpha Q}
\begin{pmatrix}
	q\\
	p
\end{pmatrix}_{t_0^-}\label{eq_flow}\\
\begin{pmatrix}
	Q\\
	P
\end{pmatrix}_{t_0^+}
&=
\begin{pmatrix}
Q\\
P-\alpha A(q,p)
\end{pmatrix}_{t_0^-}\label{eq_record}
\end{align}
\end{subequations}
where $\Phi^A_{\tau}$ is the transformation on the system's phase space that implements flowing for a ``time'' $\tau$ under the Hamiltonian flow generated by $A$. Having initialized the apparatus to its ready state~\eqref{eq_canonicalDist} prior to the interaction, then, in view of~\eqref{eq_record}, after the interaction one's state of knowledge of the apparatus, conditional on a given state of the system at the time of measurement, is
\begin{align}
	\rho(Q,P|q,p)dQdP&=\frac{\beta\Omega}{2\pi}e^{-\frac{\beta\Omega^2}{2}Q^2-\frac{\beta}{2}\left(P+\alpha A(q,p)\right)^2}dQdP.\label{eq_postInt}
\end{align}
Note that the dependence on $(q,p)$ is only through $A(q,p)$.

The trap on the apparatus is released at the moment of measurement ($\Pi(t)=0$ for $t>t_0$), so that the apparatus Hamiltonian returns to its internal setting $H_\text{app}^\text{own}$. By construction the pointer $P$ is in involution with this Hamiltonian, so it constitutes a stable record of the measurement. At this time (i.e.~any time after $t_0$) one reads the pointer on the apparatus, yielding the definite value $P^*\triangleq P(t_0^+)$, or equivalently
\begin{equation}\label{eq_recordA}
	A^*\triangleq-\frac{P^*}{\alpha}.
\end{equation}
($A^*$ is just the reading on the pointer with the scale set appropriately.) Note that this does not mean that the value of $A$ at the time of measurement is $A^*$! Rather, given this datum, the likelihood function for the value of $A$ at the time of measurement is, from~\eqref{eq_postInt},
\begin{align}\label{eq_likelihood}
		\rho(A^*|A)dA^*=\sqrt{\frac{\alpha^2\beta}{2\pi}}\exp{-\frac{\alpha^2\beta}{2}\left(A^*-A\right)^2}dA^*.
\end{align}
This completes the model of measurement. The \emph{measurement record} $A^*$, or equivalently the likelihood function~\eqref{eq_likelihood} (with $A^*$ specified), constitutes the outcome of the measurement.

\subsection{Consuming the measurement}\label{sec_consumingMeasurement}

There are two operations that one, as a recipient, should perform to consume the information of the measurement. The first is triggered by the information that the observable $A$ of the system was measured at time $t_0$ by the stipulated procedure, with specified settings $(\alpha,\beta,\Omega)$. As seen in~\eqref{eq_flow}, the interaction involved in this measurement affects the state of the system by causing it to move along the flow generated by $A$ for some unknown ``time'' $\alpha Q$. If one knew the value of $Q$ then one should change their probability distribution about the state of the system at time $t_0$ according to
\[
	\rho(q,p;t_0^-)\mapsto\rho(q,p;t_0^+)=\left[\left(\Phi^{A}_{\alpha Q}\right)_*\rho\right](q,p;t_0^-),
\]
where $(\Phi^{A}_{\tau})_*$ denotes the push-forward of the transformation $\Phi^{A}_{\tau}$, defined as $[(\Phi^{A}_\tau)_*\rho](q,p)\triangleq\rho(\Phi^{A}_{-\tau}(q,p))$; and the $+/-$ superscripts on $t_0$ are meant as a reminder that this update reflects a physical transition of the system that took place in a short time interval around $t_0$. But one does not know the value of $Q$ (see discussion in Section~\ref{sec_shiftySplit}); all that is know about it is expressed by the probability distribution~\eqref{eq_canonicalDist}. One folds this in by marginalizing over $Q$:
\begin{subequations}
\begin{equation}\label{eq_marginalize}
	\rho(q,p;t_0^+)=\sqrt{\frac{\beta\Omega^2}{2\pi}}\int_{-\infty}^\infty dQ\,e^{-\frac{\beta\Omega^2}{2}Q^2}\left[\left(\Phi^{A}_{\alpha Q}\right)_*\rho\right](q,p;t_0^-).
\end{equation}
The second operation is triggered by the information of the measurement outcome~\eqref{eq_likelihood}. One assimilates this by performing the Bayesian update $\rho_\text{pri}(q,p;t_0)\mapsto\rho_\text{post}(q,p;t_0)$,
with
\begin{align}
	\rho&_\text{post}(q,p;t_0)\propto\rho_\text{pri}(q,p;t_0)\rho(A^*|A(q,p))\notag\\
	&\propto \rho_\text{pri}(q,p;t_0)\exp{-\frac{\alpha^2\beta}{2}\left(A^*-A(q,p)\right)^2},\label{eq_BayesianUpdate}
\end{align}
\end{subequations}
where the omitted factor of proportionality is just the normalization, obtained by integrating the expression shown over the system's phase space ($\int d^nqd^np$). Since multiplication by a function of $A$ commutes with the push-forward $\left(\Phi^{A}_{\tau}\right)_*$, operations~(\ref{eq_marginalize},~\ref{eq_BayesianUpdate}) can be performed in either order to the same net effect. If~\eqref{eq_BayesianUpdate} is performed first, it corresponds to updating one's knowledge about the state the system was in before the measurement was made (i.e.~at $t_0^-)$; if second, about the state the system was left in by the measurement. Notice that if only the fact of the measurement is revealed but not the outcome (in this case it is said the outcome was \emph{discarded}), then one should only perform operation~\eqref{eq_marginalize}, not~\eqref{eq_BayesianUpdate}.

Finally, if a single number is desired as an objective quantification of the measured observable (i.e.~not biased by anyone's prior), the maximum-likelihood estimate can be given, from~\eqref{eq_likelihood}:
\begin{equation}\label{eq_outcome}
	A(t_0)\sim A^*\pm\frac{1}{\sqrt{\alpha^2\beta}}
\end{equation}
(mean $\pm$ standard deviation).\footnote{The justification for calling this number a ``mean $\pm$ standard deviation'' is that that is what it corresponds to in the posterior~\eqref{eq_BayesianUpdate} when the marginalized prior $\rho_\text{prior}(A';t_0)\triangleq\int d^nqd^np\,\delta (A'-A(q,p))\rho _\text  {prior}(q,p;t_0)$ is sufficiently flat.} I will refer to
\begin{equation}\label{eq_imprecision}
	\epsilon_A\triangleq\frac{1}{\sqrt{\alpha^2\beta}}
\end{equation}
as the \emph{imprecision} of the measurement. (But notice that to translate this to an uncertainty in a given agent's knowledge of $A$ one must first combine the likelihood function with the agent's prior, as in~\eqref{eq_BayesianUpdate}.)

\subsection{Measurement strength and apparatus quality parametrize the model}\label{sec_strengthAndQuality}

Of the three parameters $(\alpha,\beta,\Omega)$ entering the model---respectively the constant of proportionality in the interaction Hamiltonian~\eqref{eq_interactionH}, the (inverse) temperature of the apparatus, and the frequency of oscillation of the apparatus' probe around its trapped ground state---, only the two combinations
\begin{equation}\label{eq_strengthAndQuality}
	k\triangleq\frac{\alpha^2\beta}{8}\geq0\quad\text{and}\quad \qbar\triangleq\frac{2}{\beta\Omega}>0
\end{equation}
appear independently in the final results,~(\ref{eq_marginalize},~\ref{eq_BayesianUpdate}), which can be written as
\begin{subequations}
\begin{gather}
	\rho(q,p;t_0^+)=\int_{-\infty}^\infty \frac{d\tau}{\sqrt{4\pi k\qbar^2}}\,e^{-\frac{\tau^2}{4k\qbar^2}}\left[\left(\Phi^{A}_{\tau}\right)_*\rho\right](q,p;t_0^-),\label{eq_marginalize2}\\
	\rho_\text{post}(q,p;t_0)\propto\rho_\text{pri}(q,p;t_0)e^{-4k\left(A^*-A(q,p)\right)^2}.\label{eq_BayesianUpdate2}
\end{gather}
\end{subequations}
I will refer to $k$ as the \emph{strength} of the measurement; indeed, in view of~\eqref{eq_imprecision}, the larger $k$ the higher the measurement's precision.\footnote{Given the definition of $A^*$ in~\eqref{eq_recordA}, one might worry that not just $\epsilon_A$, but also $A^*$ scales with $1/\sqrt{k}$, but that's not the case. Notice, from~\eqref{eq_postInt}, that $P^*$ is drawn from a Gaussian centered at $-\alpha A$. Obviously the value of $A$ is independent of one's decision to measure it, and a fortiori of one's setting of $\alpha$. Hence it's the reading on the dial, $P^*$, that scales with $\alpha$, so $A^*$ is unaffected, in expectation, by the strength of the measurement.} Its physical dimensions are $[k]=[A]^{-2}$. For a reason to be seen next, I will refer to $\qbar$ (``q-bar'') as the \emph{inverse quality} of the apparatus (i.e.~lower values of $\qbar$ will correspond to higher-quality devices). Note that it is strictly positive and has physical dimensions of action.

\section{A Heisenberg-like precision-disturbance relation in classical physics}\label{sec_precDistRel}

The measurement model developed above is characterized by the pair $(k,\qbar)$; respectively the strength of the measurement, and the (inverse) quality of the apparatus. As has just been said, one can make the measurement of $A$ more precise by cranking up the strength, $k$, which one might think of as a knob on the experimental setup. However, notice that the more precisely $A$ is measured (i.e.~the larger $k$), the more uncertain one is about the magnitude of the back-action, or observer effect, in~\eqref{eq_marginalize2} (i.e.~the larger the variance, $2k\qbar^2$, in the ``flow time'', $\tau$). It's worth emphasizing that this observer effect is not arbitrary, but has the form of time-evolution along the Hamiltonian flow generated by the measured observable, $A$; the only thing uncertain is how much ``time'' the system flowed. One can see that this disturbance will affect some observables of the system more than others: In particular, any observable $B$ in involution with $A$ will emerge undisturbed in the immediate aftermath of the measurement.

Concretely, I find that the imprecision of a measurement~\eqref{eq_imprecision}, and the magnitude of the \emph{disturbance} caused by the measurement upon the system, 
\begin{equation}\label{eq_disturbance}
	\eta_A\triangleq \sqrt{2k\qbar^2},
\end{equation}
obey the inverse relation
\begin{equation}
	\epsilon_A\eta_A=\frac{\qbar}{2},\label{eq_precDistRel}
\end{equation}
which is fixed for a given apparatus quality; independent of the identity of the system measured, of that of the system used as measuring apparatus, of the measurement strength and of the choice of observable measured. The product on the left-hand side can easily be made larger but not smaller, as far as I can tell, suggesting that relation~\eqref{eq_precDistRel} points to a general result. (See discussions in Sections~\ref{sec_quantAndClass} and~\ref{seq_resolutionViolations}.) This Heisenberg-like precision-disturbance relation suggests an obstruction to how close one can come in a world governed by Hamiltonian mechanics to the idealization of measurement without disturbance. Note that this relation is softer than the Heisenberg uncertainty principle of quantum mechanics: For any given apparatus one will have a finite obstruction on the right-hand side of~\eqref{eq_precDistRel}, but one can always endeavor to make the obstruction smaller by cooling the apparatus further or tightening the trap (i.e.~improving apparatus quality). Instead this obstruction is of a kind with the third law, to which it is clearly related: It suggests that it is impossible by any procedure, no matter how idealized, to reduce the observer effect of measurement to zero in a finite number of operations.

\section{Continuous measurement over time, and simultaneous measurement of multiple observables}\label{sec_continuousMeasurement}

Extracting information about the system by measurement increases one's knowledge about some aspect of it. However, it has just been seen that any such measurement according to the present model will disturb the system; and this decreases one's knowledge about some other aspect of the system. For a single measurement this tradeoff is expressed by the precision-disturbance relation~\eqref{eq_precDistRel}, or in more detail by the updates~(\ref{eq_marginalize2},~\ref{eq_BayesianUpdate2}). In this section I explore the compound effect of such tradeoff due to multiple measurements; specifically, a continuous succession of vanishingly-weak measurements. (All other combinations can be recovered from this case.) This will allow me, in Section~\ref{sec_simultMeas}, to treat the cases of simultaneous measurement of multiple observables, and of inefficient measurements. The method of analysis I follow is drawn from the field of continuous quantum measurement, which addresses the corresponding problem in that setting. (See for example~\cite{jacobs2006straightforward}.)

Subdivide a finite interval of time $[0,T]$ into $N$ equal subintervals demarcated by $t_0=0<t_1<t_2<\dots<t_N=T$, with $t_j=j\Delta t$. For each $j\in\{1,\dots,N\}$, select an observable $A_j=A_j(q,p)$ of the system, and prepare for it a measurement $(k_j\Delta t,\qbar_j)$ to be carried out at time $t_j$. Notice that I have scaled the strength according to the size of the subintervals; smaller $\Delta t$ means each individual measurement is weaker, but a greater number of them fit into $[0,T]$. It will be seen that this is the right scaling for the effects to converge when the limit of smaller and smaller $\Delta t$ is taken. (Note that this changes the physical dimensions of $k_j$; they are now $[k_j]=[A_j]^{-2}\cdot\text{time}^{-1}$.) The resulting tuple of pointer readings $\mathbf{A}^*\triangleq(A^*_1,A^*_2,\dots,A^*_N)$ constitutes the measurement record for the entire succession of measurements. To assimilate the $j$-th measurement one performs the two operations~(\ref{eq_marginalize2},~\ref{eq_BayesianUpdate2}), resulting in the update
\begin{align}
	\rho&(q,p;t_{j+1})\propto \exp{-4 k_j\Delta t\left(A_j^*-A_j(q,p)\right)^2}\notag\\
	&\times\int_{-\infty}^\infty \frac{d\tau\,e^{-\frac{\tau^2}{4(k_j\Delta t)\qbar_j^2}}}{\sqrt{4\pi (k_j\Delta t)\qbar_j^2}}\left[\left(\Phi^{A_j}_{\tau}\right)_*\rho\right](q,p;t_j).\label{eq_longEq}
\end{align}
As $\Delta t$ becomes small, the exponential inside the integral vanishes except for small $\tau$. For small $\tau$, the push-forward
\begin{equation}
	\left[\left(\Phi^{A}_{\tau}\right)_*\rho\right](q_0,p_0)=\rho\left(\Phi^{A}_{-\tau}(q_0,p_0)\right)=\rho(q(-\tau),p(-\tau))
\end{equation}
can be calculated by Taylor-expanding the function $\tau\mapsto \rho(q(-\tau),p(-\tau))$ around $\tau=0$; using the chain rule to pass all time-derivatives onto $q,p$, and calculating the latter from Hamilton's equations with Hamiltonian $A$. The result is 
\begin{equation}
	\left(\Phi^{A}_{\tau}\right)_*\rho=\rho+\tau\{A,\rho\}+\frac{\tau^2}{2}\{A,\{A,\rho\}\}+\mathcal{O}(\tau^3).
\end{equation}
Putting this into~\eqref{eq_longEq}, the integral can then be done order-by-order. The odd-order terms all vanish by symmetry, leaving
\begin{align}
	\rho&(q,p;t_{j+1})\propto \exp{-4k_j\Delta t\left(A_j^*-A_j\right)^2}\notag\\
	&\quad\times\bigg(\rho+k_j\qbar_j^2\Delta t\{A_j,\{A_j,\rho\}\}+\mathcal{O}(\Delta t^2)\bigg)\Bigg|_{(q,p;t_j)}.\label{eq_longEq2}
\end{align}

\subsection{Discarded measurement record}\label{sec_discardedRec}

\begin{figure*}
	\centering
   	\includegraphics[width=\linewidth]{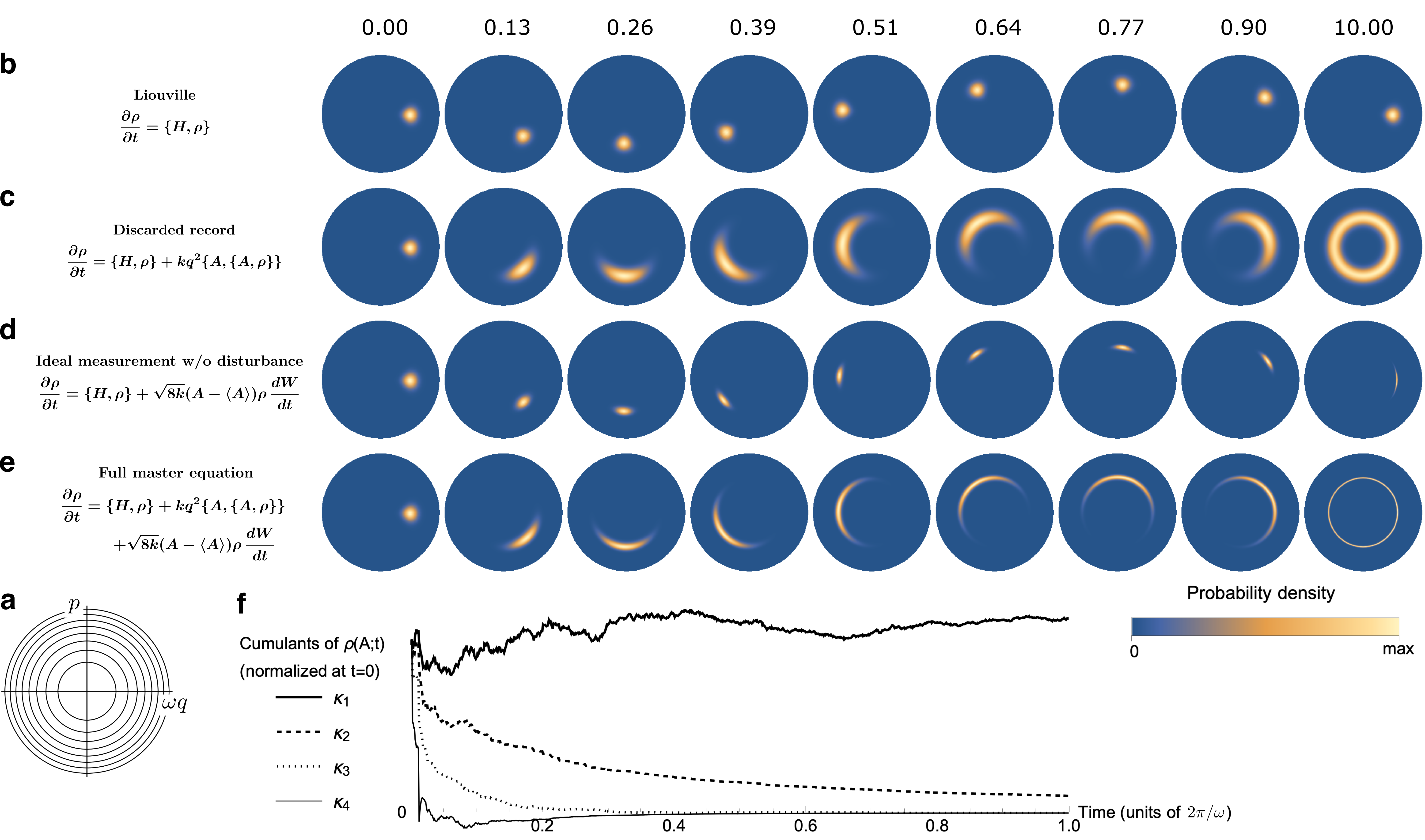}
	\caption{\label{fig_sho_master_equations}\textbf{Master equation dynamics in various measurement regimes.} Evolution of the state of knowledge $\rho(q,p;t)$ of a rational agent under master equation~\eqref{eq_ModdedStochasticLiouville} is illustrated in a simple example: The system under measurement is a 1D simple harmonic oscillator (s.h.o.); the measurement is characterized by constant $k,\qbar$ and fixed $A$; the measured observable is the energy $A=H\triangleq\frac{1}{2}(\omega^2q^2+p^2)$; and the initial distribution over phase space is unimodal. Although not proven here, three timescales are involved: that of internal dynamics, $\tau_\text{dyn}\sim1/\omega$; that of diffusion due to observer effect, $\tau_\text{dif}\sim1/k\qbar^2\omega^2$; and that of collapse due to Bayesian update on the measurement record, $\tau_\text{col}\sim1/k\Delta E^2$, where $\Delta E$ is the target certainty on $H$ (i.e.~$\tau_\text{col}$ is the characteristic timescale for the variance of $\rho(H;t)$ to fall below $\Delta E^2$). \textbf{(a)} Phase portrait showing level sets of the s.h.o.~Hamiltonian. \textbf{(b--e)} Snapshots of $\rho(q,p;t)$ at successive times, indicated at top in units of the s.h.o.~period, for four different measurement regimes (rows). The simplified master equation in each regime is indicated at left. For ease of visualization the color scheme (bottom right) is normalized anew for each plot. \textbf{(b)} Regime $\tau_\text{dyn}\ll\tau_\text{dif}, \tau_\text{col}$; describes an isolated system;~\eqref{eq_ModdedStochasticLiouville} reduces to the Liouville equation $\partial\rho/\partial t=\{H,\rho\}$. \textbf{(c)} Regime $\tau_\text{dyn}\sim\tau_\text{dif}\ll\tau_\text{col}$; describes case of discarded measurement record;~\eqref{eq_ModdedStochasticLiouville} reduces to~\eqref{eq_ModdedLiouville}. Notice entropy increase, in accordance with~\eqref{eq_Htheorem}, due to diffusion along the flow generated by $A$. \textbf{(d)} Regime $\tau_\text{dyn}\sim\tau_\text{col}\ll\tau_\text{dif}$; describes an approximation to ideal classical measurement with minimal disturbance. Notice the trend of decreasing entropy, in accordance with~\eqref{eq_antiHtheorem}, due to collapse towards the measurement outcome. \textbf{(e)} Regime $\tau_\text{dyn}\sim\tau_\text{col}\sim\tau_\text{dif}$; describes the three processes (dynamics, diffusion and collapse) happening together. Notice the tradeoff between information about $A$ and information about the conjugate quantity (s.h.o.~phase). \textbf{(f)} Evolution of the first four cumulants of $\rho(A;t)$ in regime \textbf{d} (equivalently regime \textbf{e}). For ease of visualization each cumulant is rescaled to 1 at $t=0$. Note qualitative agreement with~\eqref{eq_hierarchyOfCumulants}.}
\end{figure*}

I pause to consider the case in which the measurement record $\mathbf A^*$ is discarded. In this case one should skip update~\eqref{eq_BayesianUpdate2}, which amounts to dropping the exponential factor and the omitted proportionality factor in~\eqref{eq_longEq2}. Taking then the limit $\Delta t\to dt$ describing a continuous succession of vanishingly-weak measurements, I arrive in this case (discarded measurement record) at
\begin{align}
	\frac{\partial\rho}{\partial t}&=
	\underbrace{\{H,\rho\}}_{\substack{\text{Internal dynamics}\\ \text{Hamiltonian flow}\\ \text{Info.~preserved}}}
	+\underbrace{k\qbar^2\{A,\{A,\rho\}\}}_{\substack{\text{Disturbance}\\ \text{Diffusion along flow $\Phi^A_\tau$}\\ \text{Info.~of compatible observs.~preserved}\\ \text{Info.~of incompatible observs.~lost}}}\label{eq_ModdedLiouville}
\end{align}
where I have introduced the well-known Liouville term $\{H,\rho\}$ accounting for the internal dynamics of the system under $H$~\cite{kardar2007statistical}, which I had been ignoring until now; and all quantities shown may be explicit functions of time. This is a Liouville-like master equation, with an additional second-order term due to the disturbance caused by measurement. One can get some sense for the effect of this new term as follows. Let $B(q,p;t)$ denote any function over phase space, possibly explicitly time-dependent. Here and throughout I will use $\langle\,\cdot\,\rangle$ to denote the phase-space average:
\begin{equation}\label{eq_phaseSpaceAve}
	\langle B\rangle\triangleq\int d^nqd^np\,\rho(q,p;t)B(q,p;t).
\end{equation}
In Appendix~\ref{app_evolutionOfMean} I prove that under master equation~\eqref{eq_ModdedLiouville} any such phase-space average evolves as
\begin{align}
	\frac{d}{dt}\langle B\rangle&=\langle\lbrace B,H\rbrace\rangle+\left\langle\frac{\partial B}{\partial t}\right\rangle-k\qbar^2\left\langle\lbrace A,\log\rho\rbrace\lbrace A,B\rbrace\right\rangle.\label{eq_evolutionOfMean}
\end{align}
The first term on the right-hand side of this equation is due to the Liouville term in~\eqref{eq_ModdedLiouville}; the second term is due to any explicit time-dependence of $B$; and the third term is due to the second-order term in~\eqref{eq_ModdedLiouville}. As a special application of this equation consider $B=-\log\rho$, in which case the phase-space average is the Gibbs entropy: 
\begin{equation}\label{eq_Gibbs}
	S(t)\triangleq\langle-\log\rho\rangle.
\end{equation}
It is not hard to show that the first two terms on the right-hand side of~\eqref{eq_evolutionOfMean} vanish in this case.\footnote{Proof: By identity~\eqref{eq_cyclicIdentity} from Appendix~\ref{app_evolutionOfMean}, the first term on the right-hand side of~\eqref{eq_evolutionOfMean} can be written as $\int H\{\rho ,-\log \rho \}$, which is zero because the bracket vanishes. The second term on the right-hand side of~\eqref{eq_evolutionOfMean} is $\int \rho\frac{\partial}{\partial t}(-\log\rho)=-\int\frac{\partial \rho }{\partial t}=-\frac{d}{dt}\int\rho=-\frac{d}{dt}1=0$.} Thus I find that under dynamics~\eqref{eq_ModdedLiouville},\footnote{\eqref{eq_Htheorem} is a special case of a more general result,
\[
\frac{d}{dt}\int f(\rho)=-k\qbar^2\int f''(\rho)\{A,\rho\}^2,
\]
which holds under dynamics~\eqref{eq_ModdedLiouville}. (I omit this equation's proof, which involves steps similar to those leading to~\eqref{eq_Htheorem}.) Here $f:\mathbb R\to\mathbb R$ is any smooth function for which the shown integrals converge. It follows that for every such function which is downward-concave one has an H-theorem: $\frac{d}{dt}\int f(\rho)\geq 0$.~\eqref{eq_Htheorem} is the case $f(\rho)=-\rho\log\rho$.
}
\begin{align}
	\dot S&=k\qbar^2\left\langle\{A,\log\rho\}^2\right\rangle\geq0.\label{eq_Htheorem}
\end{align}
It is a well-known result that $S(t)$ remains constant ($\dot S=0$) under the Liouville equation $\partial\rho/\partial t=\{H,\rho\}$ (see example in Figure~\ref{fig_sho_master_equations}b). In breaking with that, it has just been found that entropy generally increases over time under~\eqref{eq_ModdedLiouville} on account of the new term. Thus the Liouville term preserves information, while the disturbance term causes information loss. Indeed, in accordance with the discussion in Section~\ref{sec_precDistRel} concerning the nature of the disturbance, this term describes diffusion along the flow lines generated by the instantaneous observable $A(q,p;t)$ (see example in Figure~\ref{fig_sho_master_equations}c). This diffusion preserves, instant-to-instant, information pertaining to observables in involution with $A(q,p;t)$, while it erases information pertaining to observables not in involution with it.

It should be noted that master equation~\eqref{eq_ModdedLiouville} has appeared in the literature before, outside the context of measurement. It appeared in~\cite{bismut1981mecanique}, which studied stochastic optimization problems. And a generalization of it appeared in~\cite{parrondo1990geometrical}, which studied Hamiltonian systems driven by colored noise.\footnote{I thank an anonymous referee for pointing out these connections.}

\subsection{Simulated measurement record}

Returning now to~\eqref{eq_longEq2}, suppose instead that the measurement record is not discarded but that one has only yet read up to the $(j-1)$-th entry; i.e.~$A^*_1$ through $A^*_{j-1}$ are known while $A^*_j$ onward are not. One would like to simulate ahead of time (say, on a computer) how one's state of knowledge will evolve as one continues to read more of the record. However, without the benefit of hindsight the upcoming record entries appear as random variables. The language for this kind of simulation is stochastic calculus. (See tutorial on stochastic calculus in~\cite{jacobs2006straightforward}.) I first ask: What should be one's probability distribution for the upcoming outcome, $A_j^*$? Making use of the likelihood function~\eqref{eq_likelihood}, this question can be answered in terms of one's current knowledge of the value of $A_j$:
\begin{align}
	\rho&(A_j^*;t_j)=\int dA_j\,\rho(A_j;t_j)\rho(A_j^*|A_j)\notag\\
	&\propto\int dA_j\,\rho(A_j;t_j)\exp{-4k_j\Delta t\left(A_j^*-A_j\right)^2}.
\end{align}
As $\Delta t\to dt$, the exponential in this expression becomes very wide and spread out as a function of $A_j$. The distribution $\rho(A_j;t_j)$ becomes very narrow by comparison, and can be replaced by a Dirac delta, which must be centered at $\langle A_j\rangle$ for the means to match. Using the delta to do the integral over $A_j$ one has, up to a normalization factor,
\begin{align}
	\rho(A_j^*;t_j)&\underset{\Delta t\to dt}{\longrightarrow}\exp{-4k_j\Delta t\left(A_j^*-\langle A_j\rangle\right)^2}.\label{eq_PDFoverPj}
\end{align}
By a simple change of variables I introduce $\Delta W_j$, one's probability distribution of which is a zero-mean Gaussian with variance $\Delta t$, and in terms of which 
\begin{equation}\label{eq_finiteStochProc}
	A_j^*=\langle A_j\rangle+\frac{1}{\sqrt{8k_j}}\frac{\Delta W_j}{\Delta t}.
\end{equation}
The value of expressing~\eqref{eq_PDFoverPj} this way is two-fold. From a simulation standpoint, one can use a random number generator to sample $\Delta W_j$ from its Gaussian distribution, and~\eqref{eq_finiteStochProc} then tells how to convert this into a sample of $A_j^*$. And from an analysis standpoint, this expression enables a very convenient form of calculation: In the limit $\Delta t\to dt$ one writes
\begin{equation}
	A^*=\langle A\rangle+\frac{1}{\sqrt{8 k}}\frac{dW}{dt},
\end{equation}
where $W(t)\triangleq\int_0^tdW$ is a standard Wiener process, with $dW$ obeying the basic rule of It\^o calculus $dW^2=dt$. Notice that $\Delta W_j$ is statistically-independent from all quantities appearing up to time $t=t_{j}$. Using $\llangle\,\cdot\,\rrangle$ to denote averaging over the Wiener process, one haves in particular, for any function $f(\rho,A)$ of the present $\rho$ and $A$: 
\begin{equation}\label{eq_averageOverWiener}
	\llangle f(\rho,A)dW\rrangle=f(\rho,A)\llangle dW\rrangle=0.
\end{equation}
Taking stock: Given $\rho(q,p;t_j)$ for a given time $t_j$ one can use it to calculate $\langle A_j\rangle$ (as in~\eqref{eq_phaseSpaceAve}), and combine this with the output of a random number generator as in~\eqref{eq_finiteStochProc} to simulate the upcoming entry of the measurement record $A_j^*$. One can then use~\eqref{eq_longEq2} to calculate what one's updated state of knowledge $\rho(q,p;t_{j+1})$ would be upon reading that entry, and iterate the process. Analytically I proceed as follows. Substitute~\eqref{eq_finiteStochProc} into~\eqref{eq_longEq2}; expand the square in the exponent, discarding the overall factor $\exp\left\{-\Delta W_j^2/2\Delta t\right\}$ which is independent of $(q,p)$; and Taylor-expand the exponential, keeping in mind that powers of $\Delta W_j$ count for ``half an order'', to obtain
\begin{align}
	&\rho(q,p;t_{j+1})\propto\notag\\
	&\quad\bigg(1-4k_j\Delta t(A_j-\langle A_j\rangle)^2+\sqrt{8k_j}\Delta W_j(A_j-\langle A_j\rangle)\notag\\
	&\qquad+4k_j\Delta W_j^2(A_j-\langle A_j\rangle)^2+\mathcal{O}(\Delta t\,\Delta W_j)\bigg)\notag\\
	&\qquad\times\bigg(\rho+k_j\qbar_j^2\Delta t\{A_j,\{A_j,\rho\}\}+\mathcal{O}(\Delta t^2)\bigg)\Bigg|_{(q,p;t_j)}.
\end{align}
In the limit of continuous measurement $\Delta t\to dt,\Delta W_j\to dW,\Delta W_j^2\to dt$ this reduces to
\begin{align}
	\rho(q,p;t+dt)&\propto\rho+k\qbar^2\{A,\{A,\rho\}\}dt\notag\\
	&\quad+\sqrt{8k}(A-\langle A\rangle)\rho\,dW\Bigg|_{(q,p;t)},
\end{align}
where again all quantities shown may be explicit functions of time. One can check that the right-hand side is already normalized, so the omitted factor of proportionality is $1$. I arrive in this case (simulated measurement record) at\footnote{All stochastic differential equations in this paper are to be interpreted in the It\^o sense.}
\begin{align}
	\frac{\partial\rho}{\partial t}&=
	\underbrace{\{H,\rho\}}_{\substack{\text{Internal dynamics}\\ \text{Hamiltonian flow}\\ \text{Info.~preserved}}}
	+\underbrace{k\qbar^2\{A,\{A,\rho\}\}}_{\substack{\text{Disturbance}\\ \text{Diffusion along flow $\Phi^A_\tau$}\\ \text{Info.~of compatible observs.~preserved}\\ \text{Info.~of incompatible observs.~lost}}}\notag\\
	&\quad+\underbrace{\sqrt{8k}(A-\langle A\rangle)\rho\,\frac{dW}{dt}}_{\substack{\text{Bayesian update} \\\text{Collapse towards measurement outcome}\\ \text{Non-linear \& non-local}\\{\llangle\Delta\text{info}\rrangle\geq0}}},\label{eq_ModdedStochasticLiouville}
\end{align}
where again I have re-introduced the Liouville term $\{H,\rho\}$ accounting for the internal dynamics of the system. Compared to~\eqref{eq_ModdedLiouville} there is here a new stochastic term appearing, which is due to assimilation of the measurement record via Bayesian update. It is interesting to note that this term is both non-linear and non-local in $\rho$, since $\langle A\rangle$ depends on the value of $\rho$ everywhere on phase space. To get some sense for the effect of this new term, in Appendix~\ref{app_antiHtheorem} I prove that under master equation~\eqref{eq_ModdedStochasticLiouville} the Gibbs entropy~\eqref{eq_Gibbs} evolves as
\begin{align}
	\dot S&=
	\underbrace{k\qbar^2\left\langle\{A,\log\rho\}^2\right\rangle}_{\substack{\text{Disturbance}\\ \Delta\text{entropy}\geq0}}\notag\\
	&\quad\underbrace{-4k\sigma_A^2-\sqrt{8k}\left\langle(A-\langle A\rangle)\log\rho\right\rangle \frac{dW}{dt}}_{\substack{\text{Bayesian update}\\ \text{Can be positive or negative}}},\label{eq_antiHtheorem1}
\end{align}
where
\begin{equation}
	\sigma_A^2=\sigma_A(t)^2\triangleq\langle(A-\langle A\rangle)^2\rangle
\end{equation}
is the variance in one's knowledge of $A(q,p;t)$ at time $t$. The first term on the r.h.s.~of~\eqref{eq_antiHtheorem1} is familiar from~\eqref{eq_Htheorem}; it describes increasing entropy due to the disturbance caused by measurement. The remaining two terms are due to the stochastic term in~\eqref{eq_ModdedStochasticLiouville}; these two together may be positive for particular measurement outcomes, but they are non-positive on average, as can be seen by invoking~\eqref{eq_averageOverWiener}:
\begin{equation}\label{eq_antiHtheorem}
	\llangle \dot S\rrangle=\underbrace{k\qbar^2\left\langle\{A,\log\rho\}^2\right\rangle}_{\substack{\text{Disturbance}\\ \Delta\text{entropy}\geq0}}
	\underbrace{-\ \ 4k\sigma_A^2}_{\substack{\text{Bayesian update}\\ \llangle \Delta\text{entropy}\rrangle\leq0}}.
\end{equation}
Thus the stochastic term in~\eqref{eq_ModdedStochasticLiouville} leads, on average, to increasing information (see example in Figure~\ref{fig_sho_master_equations}d,e). To gain further insight into the effects of this term, suppose the measured observable is fixed $A=A(q,p)$, and consider one's PDF over this observable, $\rho(A;t)$, which is just the marginal
\begin{equation}
	\rho(A';t)\triangleq\int d^nqd^np\, \delta(A(q,p)-A')\rho(q,p;t).
\end{equation}
Let $\kappa_i$ denote the $i$-th cumulant of this distribution. In Appendix~\ref{app_hierarchyOfCumulants} I prove the following hierarchy of equations describing the contribution of the stochastic term in~\eqref{eq_ModdedStochasticLiouville} to the evolution of these cumulants:
\begin{subequations}\label{eq_hierarchyOfCumulants}
\begin{align}
	d\kappa_1&=\sqrt{8k}\,\kappa_2\, dW,\\
	d\kappa_2&=\sqrt{8k}\,\kappa_3\, dW-4k(2\kappa_2^2)dt,\\
	d\kappa_3&=\sqrt{8k}\,\kappa_4\, dW-4k(6\kappa_2\kappa_3)dt,\\
	d\kappa_4&=\sqrt{8k}\,\kappa_5\, dW-4k(8\kappa_2\kappa_4+6\kappa_3^2)dt,\\
	&\dots\notag
\end{align}
\end{subequations}
Notice in particular the trends $\llangle\dot\kappa_1\rrangle=0$, $\llangle\dot\kappa_2\rrangle\sim-\llangle\kappa_2\rrangle^2$, $\llangle\dot\kappa_3\rrangle\sim-\llangle\kappa_3\rrangle$, $\llangle\dot\kappa_4\rrangle\sim-\llangle\kappa_4\rrangle$, \dots. These trends indicate that (supposing $A$ is not explicitly time-dependent and the Liouville term does not intervene too strongly) the stochastic term in~\eqref{eq_ModdedStochasticLiouville} causes all cumulants of $\rho(A;t)$ higher than second to vanish exponentially fast, leaving $\rho(A;t)$ a Gaussian; it then causes the variance to vanish like $\sim 1/t$, while the mean jiggles around in a random walk of zero drift and volatility decaying with the variance. In the limit in which the measurement process is complete, $\rho(A;t)$ converges to a delta distribution centered at the simulation's putative true value of $A$. (See example in Figure~\ref{fig_sho_master_equations}d--f.)

\subsection{Simultaneous and inefficient measurements}\label{sec_simultMeas}

Simultaneous weak measurement of multiple observables $A_1(q,p),\dots,A_s(q,p)$, whether these are in involution or not, can be handled by letting $A(q,p;t)$ in~\eqref{eq_ModdedStochasticLiouville} switch between these observables on a fast time scale. Inefficient measurements can be handled in this way too, by sporadically (on the fast time scale) discarding some of the outcomes some of the time, thus reducing~\eqref{eq_ModdedStochasticLiouville} to~\eqref{eq_ModdedLiouville} at those times. By averaging the resulting dynamics over the fast time scale one is left with
\begin{align}
	\frac{\partial\rho}{\partial t}&=\{H,\rho\}
	+\sum_{j=1}^sk_j\qbar_j^2\{A_j,\{A_j,\rho\}\}\notag\\
	&\qquad+\sum_{j=1}^s\sqrt{8\nu_jk_j}(A_j-\langle A_j\rangle)\rho\,\frac{dW_j}{dt},\label{eq_ModdedStochasticLiouvilleMultObs}
\end{align}
where $(k_j,\qbar_j,\nu_j)$ describes the measurement setup for the $j$-th observable, and $W_j(t)\triangleq\int_0^tdW_j$ are independent Wiener processes for $j\neq j'$. Here $\nu_j\in[0,1]$ is the \emph{efficiency} of the $j$-th measurement. A perfectly efficient measurement has $\nu=1$ (as in~\eqref{eq_ModdedStochasticLiouville}), while a perfectly inefficient measurement has $\nu=0$ and corresponds to discarding the outcome (as in~\eqref{eq_ModdedLiouville}).

The analogues of~\eqref{eq_antiHtheorem1} and~\eqref{eq_antiHtheorem} for the above equation are
\begin{align}
	&\dot S=\sum_{j=1}^sk_j\qbar_j^2\left\langle\{A_j,\log\rho\}^2\right\rangle\notag\\
	&-\sum_{j=1}^s\left(4\nu_jk_j\sigma_{A_j}^2+\sqrt{8\nu_jk_j}\left\langle(A_j-\langle A_j\rangle)\log\rho\right\rangle \frac{dW_j}{dt}\right),\label{eq_antiHtheorem1MultObs}
\end{align}
and
\begin{equation}\label{eq_antiHtheoremMultObs}
	\llangle \dot S\rrangle=\sum_{j=1}^sk_j\qbar_j^2\left\langle\{A_j,\log\rho\}^2\right\rangle
	-\sum_{j=1}^s4\nu_jk_j\sigma_{A_j}^2.
\end{equation}
If all the outcomes are discarded ($\nu_j=0$ for all $j$), then one is left with
\begin{align}
	\frac{\partial\rho}{\partial t}&=\{H,\rho\}
	+\sum_{j=1}^sk_j\qbar_j^2\{A_j,\{A_j,\rho\}\},\label{eq_ModdedLiouvilleMultObs}
\end{align}
which is linear, local, and deterministic; and
\begin{align}
	\dot S&=\sum_{j=1}^sk_j\qbar_j^2\left\langle\{A_j,\log\rho\}^2\right\rangle\geq0.\label{eq_HtheoremMultObs}
\end{align}

\section{Discussion}\label{sec_discussion}

\subsection{Comparing the quantum and classical uncertainty relations}\label{sec_quantAndClass}

How does the classical precision-disturbance relation~\eqref{eq_precDistRel} compare to the Heisenberg uncertainty principle of quantum mechanics? The latter can be stated in a few different forms. I will consider the Kennard-Weyl-Robertson (KWR) form in section~\ref{sec_epistLim}, where I discuss the epistemology of classical Hamiltonian ontology. Here I consider the ``joint measurement form''~\cite{arthurs1965bstj, arthurs1988quantum, ozawa1991quantum, ishikawa1991uncertainty}, pertaining to simultaneous measurement of two observables, $A$ and $B$. When the measurements are unbiased and $A,B$ are conjugate to each other this reads:
\begin{equation}\label{eq_Heisenberg}
	\epsilon_{A}\epsilon_{B}\geq\frac{\hbar}{2},
\end{equation}
where $\epsilon_{A}$ and $\epsilon_{B}$ denote the imprecisions in the measurement of $A$ and $B$, respectively;\footnote{\label{fn_realistQM} To ease comparison between the quantum and classical cases, it's convenient to speak of quantum mechanics from a realist/hidden-variable interpretation, in which measurement outcomes are outcomes \emph{about} an underlying unknown state.} and $\hbar$ is the reduced Planck constant.

One superficial difference between~\eqref{eq_precDistRel} and~\eqref{eq_Heisenberg} is that one is an equality while the other an \emph{in}equality. However, this difference is illusory. The product on the left-hand side of~\eqref{eq_precDistRel} can easily be made larger than the right-hand side, so that for a more general class of measurements one has
\begin{equation}\label{eq_precDistRel2}
	\epsilon_A\eta_A\geq\frac{\qbar}{2}.
\end{equation}
Indeed, I have defined inefficient weak measurements in Section~\ref{sec_simultMeas}, as those with $\nu<1$ in~\eqref{eq_ModdedStochasticLiouvilleMultObs}. Such measurements will fail to saturate~\eqref{eq_precDistRel2}; the extreme case of this being when the measurement outcome is discarded ($\epsilon_A\to\infty; \eta_A$ unchanged). Another way to modify the present measurement model that fails to saturate~\eqref{eq_precDistRel2} is if the apparatus' pointer fails to be in involution with $H_\text{app}^\text{own}$, so that some amount of ``deterioration'' of the measurement record can happen between the time of the system-apparatus interaction and whenever the record is read. In the opposite direction, one might ask: Could not the bound~\eqref{eq_precDistRel2} be exceeded, say, by using a probe and pointer, $(Q,P)$, that are correlated in the apparatus ready state? To achieve the latter, one would need $(Q,P)$ to \emph{not} diagonalize the quadratic form~\eqref{eq_HappWithTrap2}; namely, instead of choosing $(Q,P)=(z_{i^*},w_{i^*})$, one would choose $(Q,P)$ related to $(z_{i^*},w_{i^*})$ by some linear canonical transformation. In fact, although not proven here, I find that this approach leads to the same precision-disturbance relation~\eqref{eq_precDistRel}; the only difference (aside from the Bayesian analysis becoming more involved) is that a systematic component is added to the observer effect. This component can be corrected given the measurement outcome $A^*$; so it doesn't count towards the disturbance $\eta_A$.\footnote{\label{fn_disturbance}Recall: $\eta_A$ is defined as one's \emph{uncertainty} in $\tau$ ($\tau$ being the ``flow time'' for which the measurement caused the system to move along $\Phi^A_\tau$).} In Section~\ref{sec_Ozawa} I will analyze a second model of measurement, a classical version of Ozawa's quantum model~\cite{ozawa1988measurement}, originally proposed as a means to violate Heisenberg's relation, and it will be found that that model too is bound by~\eqref{eq_precDistRel2}. In summary, while it is easy to do worse than~\eqref{eq_precDistRel}, the present analysis suggests that it may not be possible to do better; it suggests inequality~\eqref{eq_precDistRel2} to be a general result. As the discussion in Section~\ref{sec_Ozawa} will highlight, such a result should be understood as a bound on the possibility, in Hamiltonian mechanics, of acquiring information about one observable without relinquishing the option to acquire information about other observables.

A second difference, which remains between~\eqref{eq_Heisenberg} and~\eqref{eq_precDistRel2}, is that one involves the product of two imprecisions, while the other the product of an imprecision with a disturbance magnitude. This difference can be bridged as well. Recall that the disturbance in question amounts to flowing along $\Phi^A_\tau$ for an unknown ``time'' $\tau$ whose uncertainty is $\eta_A$. Under this flow the ``rate'' of change of any observable $B$ is as given by~\eqref{eq_eom}: $\frac{d}{d\tau} B=\{B,A\}$. In particular, if $B$ is the conjugate to $A$, so that $\{B,A\}=1$, then $B$ increases monotonically at the steady rate of $1$; and the net effect of the flow on $B$ is simply to displace its value by $\tau$. (This final step can fail if $B$ has a discontinuity somewhere; so it is important that $B$ be a bona fide observable, not just a local quantity such as the phase $\phi$ of an oscillator.) So the uncertainty in the ``flow time'', $\eta_A$, translates directly into a disturbance in the value of the conjugate observable, $B$. This places a lower bound on the imprecision, $\epsilon_B$, with which any subsequent measurement can hope to determine the original value of $B$: $\epsilon_B\geq\eta_A$, with equality holding only if the measurement of $B$ is done at full strength ($k\to\infty$). Thus one has
\begin{equation}\label{eq_precPrecRel}
	\epsilon_A\epsilon_B\geq\frac{\qbar}{2},
\end{equation}
and the parallel with~\eqref{eq_Heisenberg} becomes apparent. Historically it seems that Heisenberg's own interpretation of the uncertainty principle was as a precision-disturbance relation~\cite{heisenberg1949physical}, not very different in spirit from~\eqref{eq_precDistRel2}. And in recent years work in quantum mechanics has paid considerable attention to precision-disturbance relations~\cite{ozawa2003universally, erhart2012experimental, baek2013experimental}, with some studies finding formulas similar to~\eqref{eq_precDistRel2} but with $\hbar$ in place of $\qbar$~\cite{fujikawa2012universally, busch2013proof, dressel2014certainty}.

\begin{figure}
	\centering
	\includegraphics[width=\linewidth]{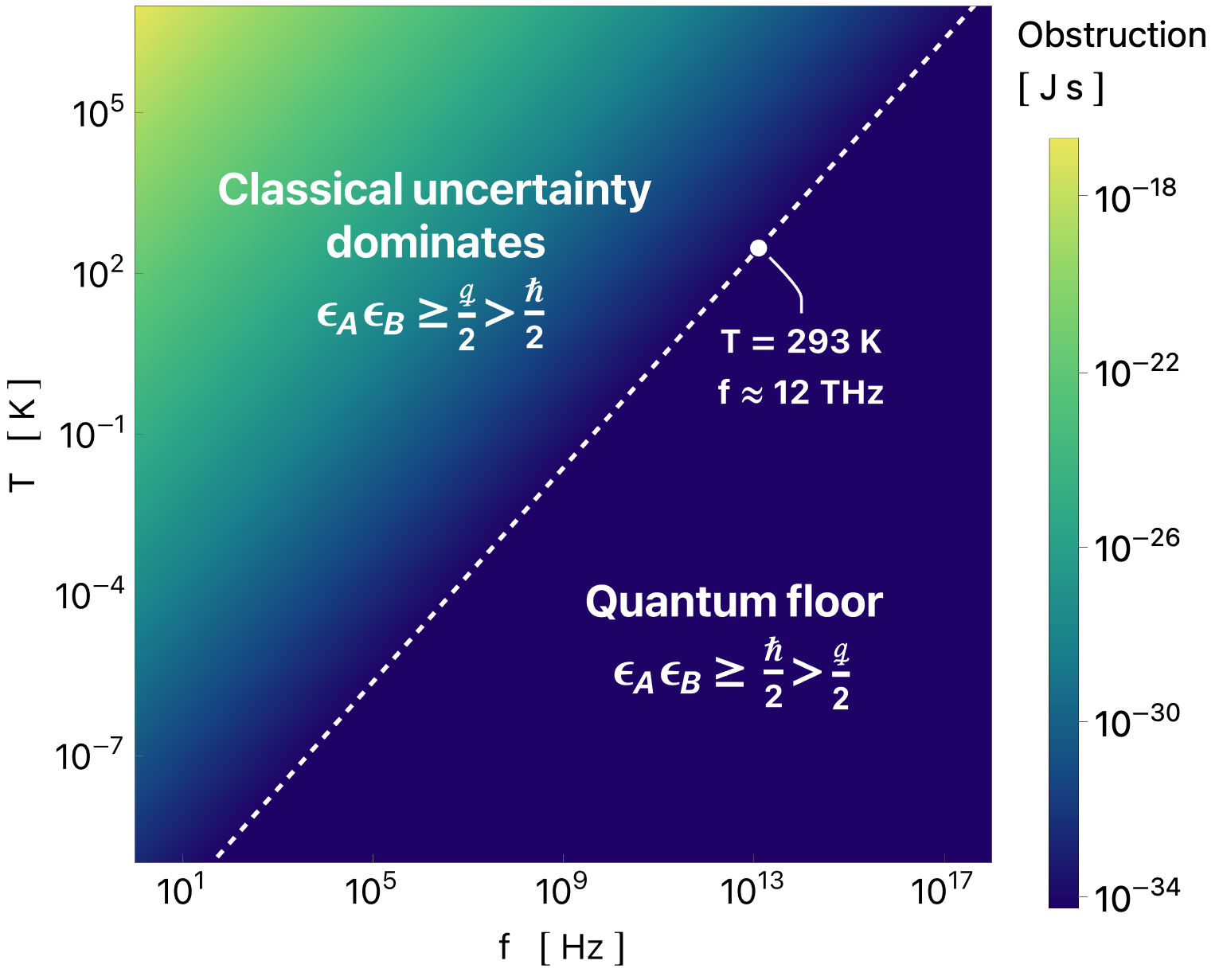} 
	\caption{\textbf{Coexistence of quantum and classical uncertainty relations in the real world.} One may expect the quantum and classical uncertainty relations to coexist in reality based on the observation that a Hamiltonian world effectively emerges from quantum mechanics at macroscopic scales. The quantum relation dominates at low apparatus temperature and/or tightly-trapped probe in the apparatus ready state; when $\qbar<\hbar$ (below the dashed diagonal). The classical relation dominates in the other direction. Here $f=\Omega/2\pi$ and $T=1/k_B\beta$. Notice that for the range of $(f,T)$ shown, the obstruction is never larger than $\sim10^{-17}$J$\cdot$s.}
	\label{fig_classical_vs_quantum}
\end{figure}
The real world is no doubt quantum mechanical, and so the Heisenberg uncertainty principle is fundamental. But as is known, as one ``zooms out'' to larger scales somehow an approximately Hamiltonian world effectively emerges (Bohr's correspondence principle and the quantum-to-classical transition). Hand in hand with the emergence of this effective Hamiltonian world I expect the classical uncertainty relation to gain traction. Figure~\ref{fig_classical_vs_quantum} illustrates how the classical and quantum relations then must coexist. For a tight enough trap and/or cold enough apparatus (below the dashed diagonal), the obstruction in~\eqref{eq_precPrecRel} is brought below $\hbar/2$ and becomes unreachable; the quantum obstruction acts like rock bottom. For less tight traps and/or warmer apparatuses the obstruction in~\eqref{eq_precPrecRel} rises above $\hbar/2$ and begins to dominate. Taken together, one may expect to have in the real world an obstruction that interpolates between these two regimes; something along the lines of
\begin{align}
	\epsilon_A\epsilon_B&\geq\ \frac{\hbar+\qbar}{2}	\ \ \text{or perhaps}\ \ 	\frac{\hbar/2}{1-e^{-\hbar/\qbar}}.	
	\label{eq_precDistRel4}
\end{align}
It will require a quantum calculation to work out the precise formula, by taking into account the probe's finite temperature and confinement in models of quantum measurement.\footnote{In this context, ``thermal'' would be a better adjective than ``classical'' for the present uncertainty relation.}  To gain some perspective for the scales involved, note from Figure~\ref{fig_classical_vs_quantum} that, at room temperature, trap frequencies any lower than about $12$THz are already enough to put one in the classical regime. At the same time, even for the highest temperatures and lowest frequencies shown in the top-left of Figure~\ref{fig_classical_vs_quantum}, the classical obstruction hardly becomes larger than $\sim10^{-17}$J$\cdot$s; an extremely small quantity by macroscopic standards. And yet, even in more moderate regimes towards the center of Figure~\ref{fig_classical_vs_quantum}, the classical obstruction may be relevant in the contexts of precision measurement, nanoengineering and molecular machines.

\subsection{On the epistemology of classical Hamiltonian ontology}\label{sec_epistLim}

Consider the KWR form of the Heisenberg uncertainty principle of quantum mechanics~\cite{griffiths2005introduction}. For a pair of conjugate observables, $A$ and $B$, it reads:
\begin{equation}\label{eq_Heisenberg1}
	\sigma_{A}\sigma_{B}\geq\frac{\hbar}{2},
\end{equation}
where $\sigma_{A}$ and $\sigma_{B}$ denote the standard deviations at a given time in one's knowledge of $A$ and $B$, respectively. (Cf.~footnote~\ref{fn_realistQM}.) This form of the uncertainty principle speaks directly to the limits of what can be known about the state of a quantum system; that is, to the epistemology of quantum ontology. In this section I ask whether the present developments allow one to establish an analogous result about the epistemology of classical Hamiltonian ontology.

Notice, first of all, the sense in which one must understand such a question. Unlike in the quantum formalism, there is nothing in the classical formalism that rules out the possibility of \emph{starting} with perfect information about conjugate observables: 
\begin{equation}\label{eq_perfectInfo}
	\rho(A',B';t)=\delta(A(t)-A')\delta(B(t)-B'),
\end{equation}
where 
\begin{align}
	\rho(A',B';t)&\triangleq\int d^nqd^np\,\delta(A(q,p)-A')\notag\\
	&\qquad\qquad\times\delta(B(q,p)-B')\rho(q,p;t).
\end{align}
Rather, the question is whether it is at all possible to \emph{arrive} at such a state of perfect information from a state of less information. In particular: Suppose one were handed a Hamiltonian system about whose state one knew nothing at all, so that $\rho$ were initially uniform on phase space. Does there exist a sequence of measurements on the system that would take $\rho$ into the perfect-information state~\eqref{eq_perfectInfo}?

Consider the direct approach of performing simultaneous measurement of $A$ and $B$, with respective measurement settings $(k_A,\qbar_A)$ and $(k_B,\qbar_B)$, and perfect efficiencies $\nu_A=\nu_B=1$. The evolution of $\rho$ is as given by master equation~\eqref{eq_ModdedStochasticLiouvilleMultObs}. Suppose that the measurements are strong enough that they come to completion on a much faster timescale than that of the system's dynamics, so that the Liouville term in~\eqref{eq_ModdedStochasticLiouvilleMultObs} can be neglected. It is a bit tricky, because one must be mindful of the rules of It\^o calculus, but one can check that, starting from an uncorrelated Gaussian distribution in $A$ and $B$ (of which the uniform distribution is the special case of infinite variances), the general solution to~\eqref{eq_ModdedStochasticLiouvilleMultObs} in this case is
\begin{equation}
	\rho(A,B;t)=\frac{1}{2\pi\sigma_{A}\sigma_{B}}\exp{-\frac{(A-\mu_{A})^2}{2\sigma_{A}^2}-\frac{(B-\mu_{B})^2}{2\sigma_{B}^2}},
\end{equation}
where the means $\mu_{A},\mu_{B}$ are stochastic functions of time evolving as
\begin{subequations}
\begin{align}
	d\mu_{A}&=\sqrt{8k_{A}}\sigma_{A}(t)^2dW_{A},\\
	d\mu_{B}&=\sqrt{8k_{B}}\sigma_{B}(t)^2dW_{B};
\end{align}
\end{subequations}
while the variances $\sigma_{A}^2,\sigma_{B}^2$ are the deterministic functions of time
\begin{subequations}\label{eq_solSigma}
\begin{align}
	\sigma_{A}(t)^2&=\frac{\qbar_B}{2}\sqrt{\frac{k_B}{k_A}}\left[\coth\left\{4\qbar_B\sqrt{k_Ak_B}(t-t_A)\right\}\right]^{l_A},\label{eq_solSigmaA}\\
	\sigma_{B}(t)^2&=\frac{\qbar_A}{2}\sqrt{\frac{k_A}{k_B}}\left[\coth\left\{4\qbar_A\sqrt{k_Ak_B}(t-t_B)\right\}\right]^{l_B},\label{eq_solSigmaB}
\end{align}
\end{subequations}
where $t_A,t_B<t$ are constants of integration, as are $l_A,l_B\in\{+1,-1\}$. Evidently, under simultaneous measurement of conjugate observables, an initially-Gaussian-uncorrelated PDF remains so for all time. Also, much like was seen in~\eqref{eq_hierarchyOfCumulants}, the mean of the distribution executes a random walk (this time in two dimensions) of volatilities proportional to the variances. However, unlike in~\eqref{eq_hierarchyOfCumulants}, now the variances converge to non-zero values as the measurements run to completion ($t\to\infty$):
\begin{equation}\label{eq_completeJointMeasurement}
	\sigma_{A}^2\to\frac{\qbar_B}{2}\sqrt{\frac{k_B}{k_A}}, \qquad	\sigma_{B}^2\to\frac{\qbar_A}{2}\sqrt{\frac{k_A}{k_B}}.
\end{equation}
This comes about because the measurement of $A$ causes collapse ``along the $A$-direction'' (along the integral curves of $\Phi^B_\tau$) and diffusion ``perpendicular to the $A$-direction'' (along the integral curves of $\Phi^A_\tau$); while the simultaneous measurement of $B$ causes the converse; and at completion of the measurement the effects precisely cancel out. Notice that~\eqref{eq_completeJointMeasurement} gives
\begin{align}
	\sigma_{A}\sigma_{B}&\to\frac{\sqrt{\qbar_A\qbar_B}}{2},
\end{align}
which begins to resemble~\eqref{eq_Heisenberg1}. Is it the case that the product $\sigma_{A}(t)\sigma_{B}(t)$ remains above this limit at all times? That depends on the exponents $l_A,l_B$. The case $l_A=+1$ gives $\sigma_A(t_A^+)\to\infty$; it describes complete ignorance about $A$ at some past time $t_A$. In contrast, the case $l_A=-1$ gives $\sigma_A(t_A^+)=0$; it describes perfect information about $A$ at the past time $t_A$. Likewise for $l_B$.\footnote{It is noteworthy that when $\sigma_A$ and $\sigma_B$ are smaller than their terminal values~\eqref{eq_completeJointMeasurement} (i.e. for $l_A=l_B=-1$) one actually \emph{loses} information by measuring! (Because the induced disturbances win over the information gains.)} Since I am interested in beginning from a state of ignorance, the relevant solution has $l_A=l_B=+1$. It then follows from~\eqref{eq_solSigma} that the inequality
\begin{align}\label{eq_uncertUncertRel}
	\sigma_{A}\sigma_{B}\geq\frac{\sqrt{\qbar_A\qbar_B}}{2}
\end{align}
holds for all times. If both measuring apparatuses are of the same (inverse) quality, $\qbar$, this further reduces to
\begin{align}\label{eq_uncertUncertRel2}
	\sigma_{A}\sigma_{B}\geq\frac{\qbar}{2}.
\end{align}
I have derived this uncertainty-uncertainty relation by considering simultaneous measurement of the pair of conjugate observables $A,B$. Could this be a general epistemic obstruction, or is there some different sequence of measurements that fares better? I leave the question open for future investigation.

\subsection{Apparatus quality as a resource}\label{sec_scaling}

I would like to suggest that what I have defined as the quality of a measuring apparatus may be thought of as a resource; as being valuable and scarce. Indeed, results~(\ref{eq_precDistRel2},~\ref{eq_precPrecRel},~\ref{eq_uncertUncertRel2}) highlight the utility of apparatus quality: At least in the present measurement model, a higher-quality apparatus means a smaller obstruction ($\qbar/2$) both to measuring without disturbance, and to simultaneous epistemic access to conjugate observables. Now I will contrast the ease with which measurement strength, $k$, can be dialed up, with the difficulty of achieving a high-quality apparatus.

Consider focusing two independent apparatuses---each capable of continuous measurement, characterized respectively by parameters $(k_1,\qbar_1)$ and $(k_2,\qbar_2)$---on the same system-observable, $A$. From~\eqref{eq_ModdedStochasticLiouvilleMultObs}, the dynamics is
\begin{align}
	\frac{\partial\rho}{\partial t}&=\{H,\rho\}
	+(k_1\qbar_1^2+k_2\qbar_2^2)\{A,\{A,\rho\}\}\notag\\
	&+\frac{\sqrt{8k_1}dW_1+\sqrt{8k_2}dW_2}{dt}(A-\langle A\rangle)\rho.
\end{align}
A sum of independent Gaussian-distributed random variables is again Gaussian-distributed, with variance equal to the sum of the variances; so one can rewrite this as
\begin{align}
	\frac{\partial\rho}{\partial t}&=\{H,\rho\}
	+k'\qbar'^2\{A,\{A,\rho\}\}
	+\sqrt{8k'}(A-\langle A\rangle)\rho\,\frac{dW'}{dt},
\end{align}
with $W'$ a standard Wiener process,
\begin{equation}
	k'\triangleq k_1+k_2,\quad\text{and}\quad\qbar'^2\triangleq\frac{k_1\qbar_1^2+k_2\qbar_2^2}{k_1+k_2}.
\end{equation}
It is clear that multiple apparatuses, each operating according to the present model, merge effectively into one also described by the model. In this the strengths add up while the (squared inverse) qualities average out. A similar thing happens when using a single apparatus and just letting it run for longer; strength compounds (Figure~\ref{fig_sho_master_equations}c--e) while quality stays fixed. The broader point is that it is straightforward to make a strong measurement out of multiple weak ones, but a high-quality apparatus can't be made out of low-quality components; it requires cold and tightly trapped probes.

\subsection{The shifty split and how to minimally bridge it: Putting present results on sound methodological ground}\label{sec_shiftySplit}

I return to a complication that was tacitly overlooked in Section~\ref{sec_measurement}. In the present measurement model, after the apparatus had interacted with the system---let me call that step the pre-measurement interaction---I stipulated that the pointer on the apparatus should be read, which would yield some definite value $P^*$. But what could it mean to ``read $P$'' if not to measure this observable of the apparatus? In the quest to define measurement, this seems to lead down an infinite regress in which the system is pre-measured by an apparatus, which must then be pre-measured, presumably by another apparatus, which must then be\dots. The passage from ``systems interacting'' to ``agent being informed'' never quite taking place. In a sense this predicament is similar to the quantum measurement problem, particularly as articulated by the Wigner's friend thought experiment~\cite{wigner1961remarks}. In both settings, the paradoxical step is the passage from what seems to be best regarded as an ontic-level description (Hamiltonian or unitary dynamics) to what seems to be best regarded as an epistemic-level operation (Bayesian updating or collapse of the wave function). At the epistemic level one speaks freely of agents, observers, measurements, observations, reading the measurement record, information, probability, Bayesian updating, collapse of the wave function. But at the ontic level all of these are complicated phenomena, often resisting satisfactory description. In lieu of such descriptions one is stuck with the \emph{shifty split}---to use a term coined by Bell~\cite{bell1990against}. 

In the present model, the shifty split was introduced at one degree of separation from the system under study: I described the system-apparatus interaction at the ontic level, then I implicitly used the following minimal postulate, \textbf{P}, as a bridge between the apparatus and agent; the latter I described at the epistemic level. Such a once-removed approach can be very useful, as exemplified by the theory of general quantum measurement~\cite{nielsen2002quantum}. 

In this section I will state postulate \textbf{P}; I will then outline what I think to be a sound methodology for constructing the epistemology of Hamiltonian mechanics; and I will point out how this methodology supports the results reached in this paper.
\begin{itemize}
	\item[\textbf{P}:] For some Hamiltonian systems and for some times, it is in principle possible for an agent to come to know (to ``read'') the exact value of one observable.
\end{itemize}
This is intended as a minimal postulate; in particular, it is agnostic as to whether the reading of an observable can be done without disturbing the system, or whether it is possible to read more than one observable of a system at a time. Starting from \textbf{P}, the epistemology of Hamiltonian mechanics should be constructed one proposition at a time. Each new candidate epistemic power should be considered in turn as a hypothesis before either being rejected or accepted. Against each new hypothesis should stand as a default the null-hypothesis: that the candidate is not in fact among the epistemic powers of an agent. In particular, and this is key: If a proposed epistemic power can neither be proven nor disproven, then it should be rejected.

Although I have not been explicit about it until now, in the present analysis I have followed this methodology. Most crucially in Section~\ref{sec_consumingMeasurement}, in connection with~\eqref{eq_marginalize}, where I did not grant the agent the power to learn new information about the original value of the apparatus' probe, $Q$, after having read the pointer, $P$. As a notable example of a proposition which does follow from this methodology: The present analysis shows that, aided by apparatuses of finite quality ($\qbar>0$), \textbf{P} implies
\begin{itemize}
	\item[\textbf{Q}:] Of any Hamiltonian system at any time, it is in principle possible for an agent to come to simultaneously know the exact values of any set of mutually-Poisson-commuting observables without disturbing them.
\end{itemize}
It is important to clearly distinguish \textbf{Q} from the stronger proposition that would drop the qualifier ``mutually-Poisson-commuting''. This stronger proposition is, of course, the commonly held idealization of classical measurement. While I have not here disproven this stronger proposition, I have also not been able to prove it. Indeed, the present analysis is able to prove it only in the case that apparatuses of infinite quality ($\qbar=0$) are available; but their availability is ruled out by the third law of thermodynamics, at least in the present model. In accordance with the above methodology, one should not (yet) accept the stronger proposition.

\subsection{Three apparent violations, two resolved}\label{seq_resolutionViolations}

\subsubsection{Ozawa's model}\label{sec_Ozawa}

In 1988 Ozawa introduced an explicit model of quantum measurement~\cite{ozawa1988measurement} which he claimed violated Heisenberg's precision-disturbance relation. The claim has been disputed and remains controversial today~\cite{busch2014colloquium,dressel2014certainty,hilgevoord2024uncertainty, ozawa2019soundness}, with the issue turning on the definitions of ``imprecision'' and ``disturbance''. In standard quantum theory observables do not generally have definite values outside of measurement, so indirect definitions for imprecision and disturbance are given in terms of pre- and post-measurement distributions. This tends to make the various authors' analyses more abstract and to obscure their interpretation. Because classical physics is free from such complications, this paper can provide a vantage point of particular clarity on the issue; which can be interesting both in the context of the quantum-theoretic controversy, as well as for its own cause: i.e.~Does the classical counterpart of Ozawa's model lead to violations of the results~(\ref{eq_precDistRel2},~\ref{eq_precPrecRel},~\ref{eq_uncertUncertRel2})?\footnote{I thank an anonymous referee for raising this question.}

Here I consider such a classical measurement model, in which the apparatus is prepared the same way as in Section~\ref{sec_readyApp}, with $P$ again as the pointer, but I replace the interaction Hamiltonian~\eqref{eq_interactionH} by Ozawa's~\cite{ozawa1988measurement}:
\begin{equation}\label{eq_interactionHOzawa}
	H_\text{int}(q,p,Q,P;t)=\frac{\pi \delta(t-t_0)}{3\sqrt{3}}\left(2(qP-pQ)-qp+QP\right).
\end{equation}
Here, for simplicity $q$ and $p$ are taken to be scalars, and $p$ is the observable being measured. Integrating Hamilton's equations, instead of~\eqref{eq_IntegrateHamsEqs} one now gets~\cite{ozawa2002position}
\begin{subequations}\label{eq_IntegrateHamsEqsOzawa}
\begin{align}
\begin{pmatrix}
	q\\
	p
\end{pmatrix}_{t_0^+}
&=
\begin{pmatrix}
	-Q\\
	p-P
\end{pmatrix}_{t_0^-}\label{eq_flowOzawa}\\
\begin{pmatrix}
	Q\\
	P
\end{pmatrix}_{t_0^+}
&=
\begin{pmatrix}
q+Q\\
p
\end{pmatrix}_{t_0^-}.\label{eq_recordOzawa}
\end{align}
\end{subequations}
As in Section~\ref{sec_measurement}, by construction the pointer $P$ constitutes a stable record of the measurement. Invoking postulate~\textbf{P} (Section~\ref{sec_shiftySplit}) one reads the pointer, which by~\eqref{eq_recordOzawa} yields the exact pre-interaction value of the desired observable: $P^*\triangleq P(t_0^+)=p(t_0^-)$. By any definition of precision concerned with pre-interaction values, this is a perfectly precise measurement: $\epsilon_p=0$.~\textcite{ozawa2002position} defines the disturbance upon the system as
\begin{align}
	\eta_\text{Ozawa}&\triangleq\sqrt{\left\langle \left(q(t_0^+)-q(t_0^-)\right)^2\right\rangle},\label{eq_OzawaDisturbance1}
\end{align}
which is the root-mean-square deviation of the conjugate observable, $q$, before and after the measuring interaction. By~\eqref{eq_flowOzawa} this equals $\sqrt{\langle(Q(t_0^-)+q(t_0^-))^2\rangle}$, where
the expectation value is over the joint (separable) prior distribution $\rho(q,p;t_0^-)\rho(Q,P;t_0^-)$, with the latter factor given by~\eqref{eq_canonicalDist}. This is a finite quantity:
\begin{align}
	\eta_\text{Ozawa}&=\sqrt{1/\beta\Omega^2+\langle q(t_0^-)^2\rangle}<\infty.\label{eq_OzawaDisturbance2}
\end{align}
This appears to be a perfectly precise measurement which does not infinitely disturb the conjugate observable. If this were indeed so, it would mean that the precision-disturbance relation~\eqref{eq_precDistRel2} is of restricted validity; merely a limitation of the model in Sec.~\ref{sec_measurement}. However, as anticipated, the apparent violation is actually a crucial matter of definitions. The notion of ``disturbance'' that is relevant to the topic of this paper---to the epistemology of Hamiltonian mechanics---is not that of deviation or displacement, but of \emph{loss in the latent information} (i.e.~loss in the possibility of access to information). In particular, since the notion of precision used here refers to the original state of the system, the relevant notion of disturbance here is of the loss in latent information regarding the original state of the system. A measurement whose net back-action on the system were to displace its state in some retraceable and correctible way would, in this sense, be a disturbance-free measurement. (Cf.~discussion in Section~\ref{sec_quantAndClass} in connection with footnote~\ref{fn_disturbance}.) Conversely, a measurement which erased all trace of the original value of some observable would, in this sense, be inflicting an infinite disturbance on the system, even if it did so by means of a finite (necessarily state-dependent) displacement. The resolution to the apparent violation is to note that Ozawa's definition~\eqref{eq_OzawaDisturbance1} is not aligned with this notion of disturbance; and Ozawa's model creates a situation in which $\eta_\text{Ozawa}$ fails to signal that all latent information on $q(t_0^-)$ gets lost. Indeed, a glance at~\eqref{eq_IntegrateHamsEqsOzawa} reveals that the measuring interaction has siphoned off from the object-system all trace of $q(t_0^-)$ and allocated it to the conjugate, $Q(t_0^+)$, of the apparatus pointer. Postulate \textbf{P}, by which I bridge the shifty split, is agnostic on the ultimate fate of $Q(t_0^+)$, but the results in this paper indicate that it is irretrievably lost in the process of reading the pointer. For example, were one to shift the shifty split one step further from the object-system, by analyzing the reading of the pointer as an exact measurement of the observable $P$ of the first apparatus, by a second apparatus of finite quality ($\qbar>0$), one would find that all trace of $Q(t_0^+)$---and hence of $q(t_0^-)$---is lost in the process.\footnote{To be more precise: Such an analysis would merely show that the measuring interaction removes the lingering trace of $q(t_0^-)$ from the conjugate of the pointer of the first apparatus ($Q$) and encodes it in the conjugate of the pointer of the second apparatus. (It is only the details of this encoding that depend on whether one uses von Neumann's interaction~\eqref{eq_interactionH} or Ozawa's~\eqref{eq_interactionHOzawa}.) The question of the ultimate availability of this information is not so much resolved as it is shifted along with the shifty split. My argument, really, as discussed in Section~\ref{sec_shiftySplit}, is that I have found no evidence to support the hypothesis that an agent can learn exactly the present value of an observable and its conjugate.} 

In summary, I find that (the classical counterpart of) Ozawa's measurement model cannot circumvent the precision-disturbance relation~\eqref{eq_precDistRel2}. Given that Ozawa's model was originally proposed as a means to violate Heisenberg's relation~\cite{ozawa1988measurement}, this finding lends substantial support to the idea that the results~(\ref{eq_precDistRel2},~\ref{eq_precPrecRel},~\ref{eq_uncertUncertRel2}) may be of general validity. Consideration of Ozawa's model has also served to sharpen the notion of disturbance that is suited to the program of this paper: as a measure not of deviation or displacement, but of loss in latent information. Relation~\eqref{eq_precDistRel2} should be understood as a bound on the possibility, in Hamiltonian mechanics, of acquiring information about one observable without relinquishing the option to acquire information about other observables.

\subsubsection{Hamiltonian bubbles; a catch-22 of epistemic access without causal access}\label{sec_bubbles} 

By application of proposition \textbf{Q} (Section~\ref{sec_shiftySplit}) at a pair of times $t_1<t_2$, to an otherwise isolated system, one can come to know a complete set of split boundary conditions for Hamilton's equations on the interval $[t_1,t_2]$; say for concreteness, all components of $q(t_1)$ and $q(t_2)$. By mathematically solving this well-posed two-point boundary-value problem, one can then know the exact state of the system, $(q(t),p(t))$, on the open interval $(t_1,t_2)$. Obviously by this method one can come to know the exact value of conjugate observables, in apparent violation of~\eqref{eq_uncertUncertRel2}. However, this is not really a contradiction. The uncertainty relation~\eqref{eq_uncertUncertRel2} refers to knowledge that can be had about the \emph{current} state of a system, while this method yields knowledge about a \emph{past} state. Indeed, notice the following stringent limitations of this method. (i)~For Hamilton's equations to hold on $(t_1,t_2)$, it is necessary that the system remain unperturbed, and hence isolated, during that interval. (ii)~Measuring the boundary conditions with the requisite perfect precision means, by~\eqref{eq_precDistRel2}, that the conjugate observables are infinitely disturbed at $t_1$ and $t_2$; so that there is no relation between $p(t_1^-)$ (the original value of $p$) and $p(t_1^+)$ (which one will come to know) nor between $p(t_2^-)$ (which one will come to know) and $p(t_2^+)$ (the final value of $p$). (iii)~Due to the linear flow of time, it is a prerequisite for completing both sets of measurements that one find oneself outside the interval $(t_1,t_2)$. Together, these restrictions have the astonishing implication that the method, which seems to grant ``forbidden'' epistemic access to degrees of freedom in excess of~\eqref{eq_uncertUncertRel2}, can do so only by forfeiting the possibility of any additional causal relationship to those degrees of freedom. It is as if the system that is undergoing such a procedure must be enclosed in a ``Hamiltonian bubble''; the boundary of which is permeable only to influences (entering the bubble at $t_1$ or exiting at $t_2$) that can be conveyed entirely by the complete set of commuting observables subject to measurement there.

\subsubsection{Filtering or postselecting from an ensemble}\label{sec_ensembles} 

Performing filtering from a large ensemble of similar systems would seem like a viable method for acquiring information about a subset of them without subjecting them to measurement---and without disturbing them. For example, by means of a slitted screen one can filter for particles with a component of their position in a narrow range. It would seem that this can be done without disturbing the conjugate momentum of the particles passing through the slit, in apparent violation of~\eqref{eq_precDistRel2}. Somewhat similarly, postselecting from an ensemble based on measurement outcomes would seem to enable epistemic access in excess of~\eqref{eq_uncertUncertRel2}. For example, consider an ensemble of systems with 2D phase spaces. Given a measuring apparatus of finite (inverse) quality $\qbar>0$ operating according to the present model, it would seem that one could perform on each system a measurement of the observable $A(q,p)\triangleq\frac12((q-q_0)^2+(p-p_0)^2)$, where the constants $q_0,p_0$ were chosen at will for each member of the ensemble. Doing each measurement at full strength so that it reveals the exact value $A^*\triangleq A(t_0)$ for its system, one's post-measurement state of knowledge for each system would then be an infinitely thin ring centered at $(q_0,p_0)$ of radius $\sqrt{2A^*}$, as in Figure~\ref{fig_sho_master_equations}e (late time). The variances in $q$ and $p$ for that post-measurement state come out to $\sigma_q^2=\sigma_p^2=A^*$, which give the product $\sigma_q\sigma_p=A^*$. This process may be thought of as if one had taken a guess that the pre-interaction state of the system was $(q,p)=(q_0,p_0)$, and then measured how far ($\sqrt{2A^*}$) that guess was from the truth. If the ensemble is large enough eventually a good enough guess would be made that $A^*<\qbar/2$, at which point one would have knowledge of $q$ and $p$ for that system exceeding the uncertainty relation~\eqref{eq_uncertUncertRel2}. Note that this would be knowledge about the \emph{post}-measurement state, so the resolution from the previous paragraph wouldn't work here.

Can these apparently legitimate methods be reconciled with the results~(\ref{eq_precDistRel2},~\ref{eq_precPrecRel},~\ref{eq_uncertUncertRel2})? I see two possibilities for doing so, not mutually exclusive. (i)~Careful analysis of these methods could reveal that they cannot, in fact, be carried out without implicitly relying on a resource (a trapped and cooled reservoir) of higher quality than $\qbar$. In the given examples: Analysis of the passage of a particle through a narrow slit could reveal that the particle's momentum \emph{is} disturbed by the thermal motion of the screen, to an extent dependent on the stiffness and temperature of the screen. Likewise, analysis of the conditions required to instantiate the interaction Hamiltonian~\eqref{eq_interactionH}, for measurement of the observable $A(q,p)\triangleq\frac12((q-q_0)^2+(p-p_0)^2)$, could reveal that fixing the ``constants'' $q_0,p_0$---which must ultimately correspond to controlled degrees of freedom of some physical system---requires a resource of higher quality than $\qbar$. (ii)~Perhaps it is sensible to define an additional kind of resource in terms of the number of systems from the ensemble that must be discarded in either method; in which case the present results may need to be extended by taking said resource into consideration together with $\qbar$. If neither of these avenues for resolution succeed then one may have a legitimate violation of~(\ref{eq_precDistRel2},~\ref{eq_precPrecRel},~\ref{eq_uncertUncertRel2}). I leave the question open for future investigation.

\subsection{Future directions}\label{sec_future}

Unexpectedly, it is now early days for studying classical measurement. The state of the theory, as developed here, seems similar to that of its quantum counterpart sometime in the 1970--80's, at least in the following respects: It is certainly after~\textcite{von2018mathematical} had introduced the first explicit model of the measurement process; but just before steam was picked up on the model-independent operational approach~\cite{davies1970operational, kraus1971general}, leading to the completely positive instrument~\cite{ozawa1984quantum, ozawa2004uncertainty}, and its stellar application to weak and continuous measurements (e.g.~\cite{belavkin1992quantum, jacobs2006straightforward}). For furthering the classical theory, inspiration is sure to be found in the many advances of its quantum-theoretic counterpart in the decades since 1970. I highlight just a few such possibilities below.

I have argued that the classical precision-disturbance relation~\eqref{eq_precDistRel2} may be of general validity in Hamiltonian mechanics. I have shown the relation to hold under a classical analogue of von Neumann's~\cite{von2018mathematical} measurement model, including extensions to weak (i.e.~continuous), inefficient and simultaneous measurements. I have also shown the relation to hold under a classical analogue of Ozawa's~\cite{ozawa1988measurement} model, originally proposed as a means to violate Heisenberg's quantum precision-disturbance relation. Further progress calls for research into alternate explicit models of classical measurement; such as the filtering and postselection approaches discussed in Section~\ref{sec_ensembles}; as well as, for example, classical analogues of the quantum models in~\cite{ozawa1990quantum}. Ultimately, what is called for is a general theory that goes beyond particular models. Towards the latter goal one might look to operational quantum theory for inspiration, particularly as it relates to continuous observables~\cite{ozawa1984quantum}. Given the predominance of continuous-valued observables in classical Hamiltonian mechanics, that approach appears particularly pertinent and instructive. It is also possible that such a goal will require delving deeper into Hamiltonian mechanics than has been done here. For instance, it may require extending Gromov's non-squeezing theorem and the notion of symplectic capacities~\cite{gromov1985pseudo, hofer1994symplectic}, from sets to probability distributions. Early attempts in this direction (which, however, only successfully deal with restricted classes of distributions evolving under quadratic Hamiltonians) can be found in~\textcite{hsiao2006fundamental, de2009symplectic}.

The present analysis of the measurement process has relied on the method of ``designer Hamiltonians''; taking for granted that the trap in~\eqref{eq_trap}, and the interaction in~\eqref{eq_interactionH} or~\eqref{eq_interactionHOzawa}, can be instantiated exactly and for free. Such assumptions may need to be revised in a more careful analysis, as briefly touched upon in Section~\ref{sec_ensembles}. These remarks point to \emph{Hamiltonian control} as a topic deserving renewed consideration, with the potential to reveal significant amendments to the present results.

Concerning continuous measurement, moving forward it will be worth honing one's intuition about the range of possible dynamics of (a rational agent's knowledge of) a system under measurement. For this it would be good to see numerical studies of~(\ref{eq_ModdedLiouville},~\ref{eq_ModdedStochasticLiouville}) applied to more interesting systems than the one-dimensional simple harmonic oscillator explored in Figure~\ref{fig_sho_master_equations}. An additional tool may be found by pursuing the classical counterpart of the theorem that converts a non-linear quantum stochastic master equation into an equivalent linear equation~\cite{wiseman1996quantum,jacobs2006straightforward}. I intend to present this theorem in an upcoming paper.

As reviewed in Section~\ref{sec_intro}, the field of information thermodynamics has developed under the almost universal assumption of ideal classical measurements (Figure~\ref{eq_idealMeasurement}). The present results, based on a more nuanced study of the measurement process, to the extent that the results are found to have a claim to generality, ought to be of input to that field; as it seems possible that amendments to that field's main results would be needed to incorporate the observer-effect of measurement.

Concerning quantum measurement, the informal discussion surrounding inequality~\eqref{eq_precDistRel4} and Figure~\ref{fig_classical_vs_quantum} suggests that, in the case that the probe is prepared in a bound thermal state, it should be possible to establish a tighter bound than that of the ``joint measurement form''~\eqref{eq_Heisenberg}~\cite{arthurs1965bstj, arthurs1988quantum, ozawa1991quantum, ishikawa1991uncertainty}. This may be an interesting direction to pursue theoretically and experimentally.

In connection with the problem of the interpretation of quantum mechanics, there is a program dating back to Einstein~\cite{einstein1935can, harrigan2010einstein} of attempting to identify and unmix a possible epistemic component of quantum theory from its ontic content. In recent times this program has made promising progress at the hands of Caves, Fuchs, and others~\cite{fuchs2002quantum, caves2002unknown, harrigan2010einstein}. In particular Spekkens~\cite{spekkens2007evidence, spekkens2016quasi}, and Bartlett, Rudolph, and Spekkens~\cite{bartlett2012reconstruction}, have illustrated how an uncircumventable epistemic limitation in an otherwise classical world, much like what is reported here, can lead to several of the phenomena usually regarded as characteristic of quantum mechanics. It will be interesting to see what these two programs can contribute to each other.

Finally I would like to pose a pair of broad, but hopefully suggestive, questions. (i)  Theoretical computer science grounds its notions of computability and complexity in concrete, if highly abstracted and idealized, physical models. If the results~(\ref{eq_precDistRel2},~\ref{eq_precPrecRel},~\ref{eq_uncertUncertRel2}) are found to have a claim to generality, how would this change theories of computation grounded in the world of classical physics?\footnote{There are parallels here to Landauer's work~\cite {landauer1961irreversibility}, establishing the thermodynamic irreversibility of certain computing processes, which launched the field of reversible computing~\cite{frank2017foundations}; and of course to quantum mechanics, which led to quantum computing.} (ii) Hamilton's equations and their underlying geometro-algebraic structure are not unique to physics; they emerge wherever the equations of a theory can be gotten out of a variational principle~\cite{arnold1990symplectic}. Indeed, in classical physics they emerge in just this way from Hamilton's principle of stationary action. In particular, optimal control theory uses essentially the same equations under the name of Pontryagin's minimum principle~\cite{kirk1970optimal}. Could the present topic have consequences for aspects of optimal control under partial information and, by extension, even for the study of intelligence? At the least, these musings illustrate the breadth of potential implications of this paper's subject.

\begin{acknowledgments}

I thank the two anonymous referees for their detailed feedback that helped improve the paper. It is a pleasure to thank Omar Eulogio L\'opez, Kurt Jacobs, and Pavel Chvykov for reading versions of the typescript and providing many useful suggestions. I'm particularly grateful to Matthew A.~Wilson, for believing in me when I most needed it and for his continued mentorship and patience. While conducting this research I was supported by the Picower Neurological Disorder Research Fund.
\end{acknowledgments}

\appendix

\section{Derivation of equation~\eqref{eq_evolutionOfMean}}\label{app_evolutionOfMean}

My objective is to derive equation~\eqref{eq_evolutionOfMean}. For brevity of notation I will omit the integration measure $d^nqd^np$ in integrals over phase space. I will make use of the identity
\begin{equation}\label{eq_cyclicIdentity}
	\int A\lbrace B,C\rbrace=\int B\lbrace C,A\rbrace=\int C\lbrace A,B\rbrace,
\end{equation}
which is valid for any smooth functions $A(q,p;t),B(q,p;t),C(q,p;t)$ as long as their product decays to zero as $\|(q,p)\|\to\infty$, so that boundary terms from integration by parts can be discarded. This identity is readily verified:
\begin{align}
	\int A\lbrace B,C\rbrace&=\int A\sum_{i=1}^n\left(\frac{\partial B}{\partial q_i}\frac{\partial C}{\partial p_i}-\frac{\partial B}{\partial p_i}\frac{\partial C}{\partial q_i}\right)\notag\\
	&=\int C\sum_{i}\left(-\frac{\partial}{\partial p_i}\left(A\frac{\partial B}{\partial q_i}\right)+\frac{\partial}{\partial q_i}\left(A\frac{\partial B}{\partial p_i}\right)\right)\notag\\
	&=\int C\sum_{i}\left(-\frac{\partial A}{\partial p_i}\frac{\partial B}{\partial q_i}+\frac{\partial A}{\partial q_i}\frac{\partial B}{\partial p_i}\right)\notag\\
	&=\int C\lbrace A,B\rbrace.
\end{align}
In my applications of the identity one of the factors will always be homogeneous in $\rho$, which it is safe to assume decays fast enough for the identity to hold (e.g.~for each $t$, $\rho(q,p;t)$ can be assumed to have compact support over phase space without any loss of physical generality.)

Now, the phase-space average of $B$ is $\langle B\rangle=\int\rho B$, and the time derivative of this is
\begin{align}
	\frac{d}{dt}\langle B\rangle=\int\left(B\frac{\partial\rho}{\partial t}+\rho\frac{\partial B}{\partial t}\right)=\int B\frac{\partial\rho}{\partial t}+\left\langle\frac{\partial B}{\partial t}\right\rangle.
\end{align}
Working with the first term on the r.h.s.~here, I substitute into it from~\eqref{eq_ModdedLiouville}:
\begin{align}
	\int B\frac{\partial\rho}{\partial t}=\int B\left(\{ H,\rho\}+k\qbar^2\{A,\{A,\rho\}\}\right).\label{eq_intermediateStep1}
\end{align}
Using identity~\eqref{eq_cyclicIdentity}, the first term on the r.h.s.~here can be written as $\int \rho\{ B,H\}=\langle\{ B,H\}\rangle$. Turning to the remaining term on the r.h.s.~of~\eqref{eq_intermediateStep1}, let $C\triangleq\{A,\rho\}$ and again use identity~\eqref{eq_cyclicIdentity}, so that the integral in this term can be written as 
\begin{align}
	\int B\{A,C\}&=\int C\{B,A\}=-\int\{A,\rho\}\{A,B\}\notag\\
	&=-\int\rho\{A,\log\rho\}\{A,B\}\notag\\
	&=-\left\langle\{A,\log\rho\}\{A,B\}\right\rangle. 
\end{align}	
All together one has
\begin{align}
	\frac{d}{dt}\langle B\rangle&=\langle\lbrace B,H\rbrace\rangle+\left\langle\frac{\partial B}{\partial t}\right\rangle-k\qbar^2\left\langle\lbrace A,\log\rho\rbrace\lbrace A,B\rbrace\right\rangle,
\end{align}
which is~\eqref{eq_evolutionOfMean}, as desired.

\section{Derivation of equation~\eqref{eq_antiHtheorem1}}\label{app_antiHtheorem}

My objective is to derive equation~\eqref{eq_antiHtheorem1}. For brevity of notation I will omit the integration measure $d^nqd^np$ in integrals over phase space. Expanding the differential of $S$ (from~\eqref{eq_Gibbs}) to second order in $d\rho$:
\begin{align}
	dS&=-\int d(\rho\log\rho)\notag\\
	&=-\int \big((\rho+d\rho)\log(\rho+d\rho)-\rho\log\rho\big)\notag\\
	&=-\int \Bigg((\rho+d\rho)\bigg(\log\rho+\frac{d\rho}{\rho}-\frac12\frac{d\rho^2}{\rho^2}\bigg)-\rho\log\rho\Bigg)\notag\\
	&=-\int \bigg((\log\rho+1)d\rho+\frac12\frac{d\rho^2}{\rho}\bigg)\notag\\
	&=-\int \bigg(\log\rho\,d\rho+\frac12\frac{d\rho^2}{\rho}\bigg).\label{eq_dSaux}
\end{align}
(In the last step I used the fact that $\int d\rho=d\int\rho=d1=0$.) I will now substitute into here for $d\rho$ from~\eqref{eq_ModdedStochasticLiouville}. However, notice that the non-stochastic terms from that equation will only contribute linearly (since terms of order $dt\,dW$ and $dt^2$ are negligible), so their final contribution to $dS$ will be the same as already deduced in connection with master equation~\eqref{eq_ModdedLiouville} (cf.~\eqref{eq_Htheorem}). One therefore need only calculate here the contribution to $dS$ of the stochastic term in~\eqref{eq_ModdedStochasticLiouville}; that is of $d\rho=\sqrt{8k}(A-\langle A\rangle)\rho\,dW$. Substituting this into~\eqref{eq_dSaux}, and in the following step using the rule of It\^o calculus $dW^2=dt$:
\begin{align}
	dS&=-\int \bigg(\log\rho\left(\sqrt{8k}(A-\langle A\rangle)\rho\,dW\right)\notag\\
	&\qquad\qquad+\frac12\frac{\left(\sqrt{8 k}(A-\langle A\rangle)\rho\,dW\right)^2}{\rho}\bigg)\notag\\
	&=-\sqrt{8 k}\,dW\int (A-\langle A\rangle)\rho\log\rho\notag\\
	&\qquad\qquad-4kdt\int(A-\langle A\rangle)^2\rho\notag\\
	&=-\sqrt{8 k}\,dW\left\langle (A-\langle A\rangle)\log\rho\right\rangle-4k\sigma_A^2dt.
\end{align}
This, together with the contribution~\eqref{eq_Htheorem} due to the non-stochastic terms from~\eqref{eq_ModdedStochasticLiouville}, gives~\eqref{eq_antiHtheorem1}, as desired.

\section{Derivation of the hierarchy of equations~\eqref{eq_hierarchyOfCumulants}}\label{app_hierarchyOfCumulants}

My objective is to derive the hierarchy of equations~\eqref{eq_hierarchyOfCumulants}, which describes the contribution of the stochastic term in~\eqref{eq_ModdedStochasticLiouville} to the evolution of the cumulants of $\rho(A;t)$ when $A=A(q,p)$ is not explicitly time-dependent. For brevity of notation I will omit the integration measure $d^nqd^np$ in integrals over phase space. Consider the cumulant-generating function for $\rho(A;t)$:
\begin{equation}\label{eq_cumulantGenFunct}
	f(z;t)\triangleq\log\left\langle e^{zA}\right\rangle\triangleq \kappa_1(t)\frac{z}{1!}+\kappa_2(t)\frac{z^2}{2!}+\kappa_3(t)\frac{z^3}{3!}+\dots.
\end{equation}
Let $df$ denote the differential of this function with respect to time, and $f'$ denote its derivative with respect to the dummy variable $z$. Expanding the differential of $f$ to second order in $d\rho$:
\begin{align}
	df&=d\left(\log\int \rho\,e^{zA}\right)=\log\int (\rho+d\rho) e^{zA}-\log\int \rho e^{zA}\notag\\
	&=\left(\frac{\int d\rho\,e^{zA}}{\int \rho\,e^{zA}}\right)-\frac{1}{2}\left(\frac{\int d\rho\,e^{zA}}{\int \rho\,e^{zA}}\right)^2.
\end{align}
Substituting into here the stochastic term from~\eqref{eq_ModdedStochasticLiouville} (that is $d\rho=\sqrt{8k}(A-\langle A\rangle)\rho\,dW$), and using the rule of It\^o calculus $dW^2=dt$:
\begin{align}
	df&=\sqrt{8k}\,dW\left(\frac{\int (A-\langle A\rangle)\rho\,e^{zA}}{\int \rho\,e^{zA}}\right)\notag\\
	&\qquad\qquad-4kdt\left(\frac{\int (A-\langle A\rangle)\rho\,e^{zA}}{\int \rho\,e^{zA}}\right)^2\notag\\
	&=\sqrt{8 k}\,dW\left(f'-\langle A\rangle\right)-4kdt\left(f'-\langle A\rangle\right)^2.
\end{align}
Writing $f$ in terms of its cumulant expansion~\eqref{eq_cumulantGenFunct}, and noting that $\langle A\rangle=\kappa_1$:
\begin{widetext}
\begin{align}
	d\kappa_1\frac{z}{1!}+d\kappa_2\frac{z^2}{2!}+d\kappa_3\frac{z^3}{3!}+\dots&=\sqrt{8k}\,dW\left(\kappa_2\frac{z}{1!}+\kappa_3\frac{z^2}{2!}+\kappa_4\frac{z^3}{3!}+\dots\right)-4kdt\left(\kappa_2\frac{z}{1!}+\kappa_3\frac{z^2}{2!}+\kappa_4\frac{z^3}{3!}+\dots\right)^2.
\end{align}
\end{widetext}
Expanding the square on the r.h.s.~and equating coefficients of corresponding powers of $z$ yields the hierarchy of equations~\eqref{eq_hierarchyOfCumulants}, as desired.

\bibliography{measurement}

\begin{thebibliography}{77}%
\makeatletter
\providecommand \@ifxundefined [1]{%
 \@ifx{#1\undefined}
}%
\providecommand \@ifnum [1]{%
 \ifnum #1\expandafter \@firstoftwo
 \else \expandafter \@secondoftwo
 \fi
}%
\providecommand \@ifx [1]{%
 \ifx #1\expandafter \@firstoftwo
 \else \expandafter \@secondoftwo
 \fi
}%
\providecommand \natexlab [1]{#1}%
\providecommand \enquote  [1]{``#1''}%
\providecommand \bibnamefont  [1]{#1}%
\providecommand \bibfnamefont [1]{#1}%
\providecommand \citenamefont [1]{#1}%
\providecommand \href@noop [0]{\@secondoftwo}%
\providecommand \href [0]{\begingroup \@sanitize@url \@href}%
\providecommand \@href[1]{\@@startlink{#1}\@@href}%
\providecommand \@@href[1]{\endgroup#1\@@endlink}%
\providecommand \@sanitize@url [0]{\catcode `\\12\catcode `\$12\catcode
  `\&12\catcode `\#12\catcode `\^12\catcode `\_12\catcode `\%12\relax}%
\providecommand \@@startlink[1]{}%
\providecommand \@@endlink[0]{}%
\providecommand \url  [0]{\begingroup\@sanitize@url \@url }%
\providecommand \@url [1]{\endgroup\@href {#1}{\urlprefix }}%
\providecommand \urlprefix  [0]{URL }%
\providecommand \Eprint [0]{\href }%
\providecommand \doibase [0]{https://doi.org/}%
\providecommand \selectlanguage [0]{\@gobble}%
\providecommand \bibinfo  [0]{\@secondoftwo}%
\providecommand \bibfield  [0]{\@secondoftwo}%
\providecommand \translation [1]{[#1]}%
\providecommand \BibitemOpen [0]{}%
\providecommand \bibitemStop [0]{}%
\providecommand \bibitemNoStop [0]{.\EOS\space}%
\providecommand \EOS [0]{\spacefactor3000\relax}%
\providecommand \BibitemShut  [1]{\csname bibitem#1\endcsname}%
\let\auto@bib@innerbib\@empty
\bibitem [{\citenamefont {Arthurs}\ and\ \citenamefont
  {Kelly}(1965)}]{arthurs1965bstj}%
  \BibitemOpen
  \bibfield  {author} {\bibinfo {author} {\bibfnamefont {E.}~\bibnamefont
  {Arthurs}}\ and\ \bibinfo {author} {\bibfnamefont {J.~L.}\ \bibnamefont
  {Kelly}},\ }\bibfield  {title} {\bibinfo {title} {{BSTJ} {B}riefs: On the
  simultaneous measurement of a pair of conjugate observables},\ }\href@noop {}
  {\bibfield  {journal} {\bibinfo  {journal} {Bell Syst. Tech. J.}\ }\textbf
  {\bibinfo {volume} {44}},\ \bibinfo {pages} {725} (\bibinfo {year}
  {1965})}\BibitemShut {NoStop}%
\bibitem [{\citenamefont {Arthurs}\ and\ \citenamefont
  {Goodman}(1988)}]{arthurs1988quantum}%
  \BibitemOpen
  \bibfield  {author} {\bibinfo {author} {\bibfnamefont {E.}~\bibnamefont
  {Arthurs}}\ and\ \bibinfo {author} {\bibfnamefont {M.~S.}\ \bibnamefont
  {Goodman}},\ }\bibfield  {title} {\bibinfo {title} {Quantum correlations: A
  generalized {H}eisenberg uncertainty relation},\ }\href@noop {} {\bibfield
  {journal} {\bibinfo  {journal} {Phys. Rev. Lett.}\ }\textbf {\bibinfo
  {volume} {60}},\ \bibinfo {pages} {2447} (\bibinfo {year}
  {1988})}\BibitemShut {NoStop}%
\bibitem [{\citenamefont {Ozawa}(1991)}]{ozawa1991quantum}%
  \BibitemOpen
  \bibfield  {author} {\bibinfo {author} {\bibfnamefont {M.}~\bibnamefont
  {Ozawa}},\ }\bibfield  {title} {\bibinfo {title} {Quantum limits of
  measurements and uncertainty principle},\ }in\ \href@noop {} {\emph {\bibinfo
  {booktitle} {Quantum Aspects of Optical Communications}}},\ \bibinfo {editor}
  {edited by\ \bibinfo {editor} {\bibfnamefont {C.}~\bibnamefont
  {Bendjaballah}}, \bibinfo {editor} {\bibfnamefont {O.}~\bibnamefont
  {Hirota}},\ and\ \bibinfo {editor} {\bibfnamefont {S.}~\bibnamefont
  {Reynaud}}}\ (\bibinfo  {publisher} {Springer},\ \bibinfo {year} {1991})\
  pp.\ \bibinfo {pages} {1--17}\BibitemShut {NoStop}%
\bibitem [{\citenamefont {Ishikawa}(1991)}]{ishikawa1991uncertainty}%
  \BibitemOpen
  \bibfield  {author} {\bibinfo {author} {\bibfnamefont {S.}~\bibnamefont
  {Ishikawa}},\ }\bibfield  {title} {\bibinfo {title} {Uncertainty relations in
  simultaneous measurements for arbitrary observables},\ }\href@noop {}
  {\bibfield  {journal} {\bibinfo  {journal} {Rep. Math. Phys.}\ }\textbf
  {\bibinfo {volume} {29}},\ \bibinfo {pages} {257} (\bibinfo {year}
  {1991})}\BibitemShut {NoStop}%
\bibitem [{\citenamefont {Heisenberg}(1949)}]{heisenberg1949physical}%
  \BibitemOpen
  \bibfield  {author} {\bibinfo {author} {\bibfnamefont {W.}~\bibnamefont
  {Heisenberg}},\ }\href@noop {} {\emph {\bibinfo {title} {The physical
  principles of the quantum theory}}}\ (\bibinfo  {publisher} {Dover},\
  \bibinfo {year} {1949})\BibitemShut {NoStop}%
\bibitem [{\citenamefont {Lamb}\ and\ \citenamefont
  {Fearn}(1996)}]{lamb1996classical}%
  \BibitemOpen
  \bibfield  {author} {\bibinfo {author} {\bibfnamefont {W.~E.}\ \bibnamefont
  {Lamb}}\ and\ \bibinfo {author} {\bibfnamefont {H.}~\bibnamefont {Fearn}},\
  }\bibfield  {title} {\bibinfo {title} {Classical theory of measurement: A big
  step towards the quantum theory of measurement},\ }in\ \href@noop {} {\emph
  {\bibinfo {booktitle} {Amazing Light}}}\ (\bibinfo  {publisher} {Springer},\
  \bibinfo {year} {1996})\ pp.\ \bibinfo {pages} {373--389}\BibitemShut
  {NoStop}%
\bibitem [{\citenamefont {Morgan}(2020)}]{morgan2020algebraic}%
  \BibitemOpen
  \bibfield  {author} {\bibinfo {author} {\bibfnamefont {P.}~\bibnamefont
  {Morgan}},\ }\bibfield  {title} {\bibinfo {title} {An algebraic approach to
  {K}oopman classical mechanics},\ }\href@noop {} {\bibfield  {journal}
  {\bibinfo  {journal} {Ann. Phys.}\ }\textbf {\bibinfo {volume} {414}},\
  \bibinfo {pages} {168090} (\bibinfo {year} {2020})}\BibitemShut {NoStop}%
\bibitem [{\citenamefont {Katagiri}(2020)}]{katagiri2020measurement}%
  \BibitemOpen
  \bibfield  {author} {\bibinfo {author} {\bibfnamefont {S.}~\bibnamefont
  {Katagiri}},\ }\bibfield  {title} {\bibinfo {title} {Measurement theory in
  classical mechanics},\ }\href@noop {} {\bibfield  {journal} {\bibinfo
  {journal} {Prog. Theor. Exp. Phys.}\ }\textbf {\bibinfo {volume} {2020}},\
  \bibinfo {pages} {063A02} (\bibinfo {year} {2020})}\BibitemShut {NoStop}%
\bibitem [{\citenamefont {Parrondo}\ \emph {et~al.}(2015)\citenamefont
  {Parrondo}, \citenamefont {Horowitz},\ and\ \citenamefont
  {Sagawa}}]{parrondo2015thermodynamics}%
  \BibitemOpen
  \bibfield  {author} {\bibinfo {author} {\bibfnamefont {J.~M.}\ \bibnamefont
  {Parrondo}}, \bibinfo {author} {\bibfnamefont {J.~M.}\ \bibnamefont
  {Horowitz}},\ and\ \bibinfo {author} {\bibfnamefont {T.}~\bibnamefont
  {Sagawa}},\ }\bibfield  {title} {\bibinfo {title} {Thermodynamics of
  information},\ }\href@noop {} {\bibfield  {journal} {\bibinfo  {journal}
  {Nat. Phys.}\ }\textbf {\bibinfo {volume} {11}},\ \bibinfo {pages} {131}
  (\bibinfo {year} {2015})}\BibitemShut {NoStop}%
\bibitem [{\citenamefont {Collier}(1990)}]{collier1990two}%
  \BibitemOpen
  \bibfield  {author} {\bibinfo {author} {\bibfnamefont {J.}~\bibnamefont
  {Collier}},\ }\bibfield  {title} {\bibinfo {title} {Two faces of {M}axwell's
  demon reveal the nature of irreversibility},\ }\href@noop {} {\bibfield
  {journal} {\bibinfo  {journal} {Stud. Hist. Philos. Sci.}\ }\textbf {\bibinfo
  {volume} {21}},\ \bibinfo {pages} {22J} (\bibinfo {year} {1990})}\BibitemShut
  {NoStop}%
\bibitem [{\citenamefont {Szilard}(1929)}]{szilard1929entropieverminderung}%
  \BibitemOpen
  \bibfield  {author} {\bibinfo {author} {\bibfnamefont {L.}~\bibnamefont
  {Szilard}},\ }\bibfield  {title} {\bibinfo {title} {{\"U}ber die
  entropieverminderung in einem thermodynamischen system bei eingriffen
  intelligenter wesen},\ }\href@noop {} {\bibfield  {journal} {\bibinfo
  {journal} {Z. Phys.}\ }\textbf {\bibinfo {volume} {53}},\ \bibinfo {pages}
  {840} (\bibinfo {year} {1929})}\BibitemShut {NoStop}%
\bibitem [{\citenamefont {von Neumann}(2018)}]{von2018mathematical}%
  \BibitemOpen
  \bibfield  {author} {\bibinfo {author} {\bibfnamefont {J.}~\bibnamefont {von
  Neumann}},\ }\href@noop {} {\emph {\bibinfo {title} {Mathematical foundations
  of quantum mechanics: New edition}}}\ (\bibinfo  {publisher} {Princeton
  university press},\ \bibinfo {year} {2018})\BibitemShut {NoStop}%
\bibitem [{\citenamefont {Brillouin}(1951)}]{brillouin1951maxwell}%
  \BibitemOpen
  \bibfield  {author} {\bibinfo {author} {\bibfnamefont {L.}~\bibnamefont
  {Brillouin}},\ }\bibfield  {title} {\bibinfo {title} {Maxwell's demon cannot
  operate: Information and entropy. {I}},\ }\href@noop {} {\bibfield  {journal}
  {\bibinfo  {journal} {J. Appl. Phys.}\ }\textbf {\bibinfo {volume} {22}},\
  \bibinfo {pages} {334} (\bibinfo {year} {1951})}\BibitemShut {NoStop}%
\bibitem [{\citenamefont {Brillouin}(1953)}]{brillouin1953negentropy}%
  \BibitemOpen
  \bibfield  {author} {\bibinfo {author} {\bibfnamefont {L.}~\bibnamefont
  {Brillouin}},\ }\bibfield  {title} {\bibinfo {title} {The negentropy
  principle of information},\ }\href@noop {} {\bibfield  {journal} {\bibinfo
  {journal} {J. Appl. Phys.}\ }\textbf {\bibinfo {volume} {24}},\ \bibinfo
  {pages} {1152} (\bibinfo {year} {1953})}\BibitemShut {NoStop}%
\bibitem [{\citenamefont {Gabor}(1961)}]{gabor1961perpetuum}%
  \BibitemOpen
  \bibfield  {author} {\bibinfo {author} {\bibfnamefont {D.}~\bibnamefont
  {Gabor}},\ }\bibfield  {title} {\bibinfo {title} {Light and information},\
  }in\ \href@noop {} {\emph {\bibinfo {booktitle} {Progress in optics}}},\
  Vol.~\bibinfo {volume} {1}\ (\bibinfo  {publisher} {Elsevier},\ \bibinfo
  {year} {1961})\ pp.\ \bibinfo {pages} {109--153}\BibitemShut {NoStop}%
\bibitem [{\citenamefont {Bennett}(1982)}]{bennett1982thermodynamics}%
  \BibitemOpen
  \bibfield  {author} {\bibinfo {author} {\bibfnamefont {C.~H.}\ \bibnamefont
  {Bennett}},\ }\bibfield  {title} {\bibinfo {title} {The thermodynamics of
  computation—a review},\ }\href@noop {} {\bibfield  {journal} {\bibinfo
  {journal} {Int. J. Theor. Phys.}\ }\textbf {\bibinfo {volume} {21}},\
  \bibinfo {pages} {905} (\bibinfo {year} {1982})}\BibitemShut {NoStop}%
\bibitem [{\citenamefont {Bennett}(1987)}]{bennett1987demons}%
  \BibitemOpen
  \bibfield  {author} {\bibinfo {author} {\bibfnamefont {C.~H.}\ \bibnamefont
  {Bennett}},\ }\bibfield  {title} {\bibinfo {title} {Demons, engines and the
  second law},\ }\href@noop {} {\bibfield  {journal} {\bibinfo  {journal} {Sci.
  Am.}\ }\textbf {\bibinfo {volume} {257}},\ \bibinfo {pages} {108} (\bibinfo
  {year} {1987})}\BibitemShut {NoStop}%
\bibitem [{\citenamefont {Landauer}(1961)}]{landauer1961irreversibility}%
  \BibitemOpen
  \bibfield  {author} {\bibinfo {author} {\bibfnamefont {R.}~\bibnamefont
  {Landauer}},\ }\bibfield  {title} {\bibinfo {title} {Irreversibility and heat
  generation in the computing process},\ }\href@noop {} {\bibfield  {journal}
  {\bibinfo  {journal} {IBM J. Res. Dev.}\ }\textbf {\bibinfo {volume} {5}},\
  \bibinfo {pages} {183} (\bibinfo {year} {1961})}\BibitemShut {NoStop}%
\bibitem [{\citenamefont {Earman}\ and\ \citenamefont
  {Norton}(1998)}]{earman1998exorcist}%
  \BibitemOpen
  \bibfield  {author} {\bibinfo {author} {\bibfnamefont {J.}~\bibnamefont
  {Earman}}\ and\ \bibinfo {author} {\bibfnamefont {J.~D.}\ \bibnamefont
  {Norton}},\ }\bibfield  {title} {\bibinfo {title} {Exorcist {XIV}: The wrath
  of {M}axwell’s demon. {P}art {I}. {F}rom {M}axwell to {S}zilard},\
  }\href@noop {} {\bibfield  {journal} {\bibinfo  {journal} {Stud. Hist. Phil.
  Mod. Phys.}\ }\textbf {\bibinfo {volume} {29}},\ \bibinfo {pages} {435}
  (\bibinfo {year} {1998})}\BibitemShut {NoStop}%
\bibitem [{\citenamefont {Earman}\ and\ \citenamefont
  {Norton}(1999)}]{earman1999exorcist}%
  \BibitemOpen
  \bibfield  {author} {\bibinfo {author} {\bibfnamefont {J.}~\bibnamefont
  {Earman}}\ and\ \bibinfo {author} {\bibfnamefont {J.~D.}\ \bibnamefont
  {Norton}},\ }\bibfield  {title} {\bibinfo {title} {Exorcist {XIV}: The wrath
  of {M}axwell’s demon. {P}art {II}. {F}rom {S}zilard to {L}andauer and
  beyond},\ }\href@noop {} {\bibfield  {journal} {\bibinfo  {journal} {Stud.
  Hist. Phil. Mod. Phys.}\ }\textbf {\bibinfo {volume} {30}},\ \bibinfo {pages}
  {1} (\bibinfo {year} {1999})}\BibitemShut {NoStop}%
\bibitem [{\citenamefont {Sagawa}\ and\ \citenamefont
  {Ueda}(2009)}]{sagawa2009minimal}%
  \BibitemOpen
  \bibfield  {author} {\bibinfo {author} {\bibfnamefont {T.}~\bibnamefont
  {Sagawa}}\ and\ \bibinfo {author} {\bibfnamefont {M.}~\bibnamefont {Ueda}},\
  }\bibfield  {title} {\bibinfo {title} {Minimal energy cost for thermodynamic
  information processing: Measurement and information erasure},\ }\href@noop {}
  {\bibfield  {journal} {\bibinfo  {journal} {Phys. Rev. Lett.}\ }\textbf
  {\bibinfo {volume} {102}},\ \bibinfo {pages} {250602} (\bibinfo {year}
  {2009})}\BibitemShut {NoStop}%
\bibitem [{\citenamefont {Sagawa}(2012)}]{sagawa2012thermodynamics}%
  \BibitemOpen
  \bibfield  {author} {\bibinfo {author} {\bibfnamefont {T.}~\bibnamefont
  {Sagawa}},\ }\bibfield  {title} {\bibinfo {title} {Thermodynamics of
  information processing in small systems},\ }\href@noop {} {\bibfield
  {journal} {\bibinfo  {journal} {Prog. Theor. Phys.}\ }\textbf {\bibinfo
  {volume} {127}},\ \bibinfo {pages} {1} (\bibinfo {year} {2012})}\BibitemShut
  {NoStop}%
\bibitem [{\citenamefont {Mandal}\ and\ \citenamefont
  {Jarzynski}(2012)}]{mandal2012work}%
  \BibitemOpen
  \bibfield  {author} {\bibinfo {author} {\bibfnamefont {D.}~\bibnamefont
  {Mandal}}\ and\ \bibinfo {author} {\bibfnamefont {C.}~\bibnamefont
  {Jarzynski}},\ }\bibfield  {title} {\bibinfo {title} {Work and information
  processing in a solvable model of {M}axwell’s demon},\ }\href@noop {}
  {\bibfield  {journal} {\bibinfo  {journal} {P. Natl. Acad. Sci. U.S.A.}\
  }\textbf {\bibinfo {volume} {109}},\ \bibinfo {pages} {11641} (\bibinfo
  {year} {2012})}\BibitemShut {NoStop}%
\bibitem [{\citenamefont {Barato}\ and\ \citenamefont
  {Seifert}(2013)}]{barato2013autonomous}%
  \BibitemOpen
  \bibfield  {author} {\bibinfo {author} {\bibfnamefont {A.~C.}\ \bibnamefont
  {Barato}}\ and\ \bibinfo {author} {\bibfnamefont {U.}~\bibnamefont
  {Seifert}},\ }\bibfield  {title} {\bibinfo {title} {An autonomous and
  reversible {M}axwell's demon},\ }\href@noop {} {\bibfield  {journal}
  {\bibinfo  {journal} {Europhys. Lett.}\ }\textbf {\bibinfo {volume} {101}},\
  \bibinfo {pages} {60001} (\bibinfo {year} {2013})}\BibitemShut {NoStop}%
\bibitem [{\citenamefont {Mandal}\ \emph {et~al.}(2013)\citenamefont {Mandal},
  \citenamefont {Quan},\ and\ \citenamefont {Jarzynski}}]{mandal2013maxwell}%
  \BibitemOpen
  \bibfield  {author} {\bibinfo {author} {\bibfnamefont {D.}~\bibnamefont
  {Mandal}}, \bibinfo {author} {\bibfnamefont {H.~T.}\ \bibnamefont {Quan}},\
  and\ \bibinfo {author} {\bibfnamefont {C.}~\bibnamefont {Jarzynski}},\
  }\bibfield  {title} {\bibinfo {title} {{M}axwell’s refrigerator: An exactly
  solvable model},\ }\href@noop {} {\bibfield  {journal} {\bibinfo  {journal}
  {Phys. Rev. Lett.}\ }\textbf {\bibinfo {volume} {111}},\ \bibinfo {pages}
  {030602} (\bibinfo {year} {2013})}\BibitemShut {NoStop}%
\bibitem [{\citenamefont {Hoppenau}\ and\ \citenamefont
  {Engel}(2014)}]{hoppenau2014energetics}%
  \BibitemOpen
  \bibfield  {author} {\bibinfo {author} {\bibfnamefont {J.}~\bibnamefont
  {Hoppenau}}\ and\ \bibinfo {author} {\bibfnamefont {A.}~\bibnamefont
  {Engel}},\ }\bibfield  {title} {\bibinfo {title} {On the energetics of
  information exchange},\ }\href@noop {} {\bibfield  {journal} {\bibinfo
  {journal} {Europhys. Lett.}\ }\textbf {\bibinfo {volume} {105}},\ \bibinfo
  {pages} {50002} (\bibinfo {year} {2014})}\BibitemShut {NoStop}%
\bibitem [{\citenamefont {Strasberg}\ \emph {et~al.}(2013)\citenamefont
  {Strasberg}, \citenamefont {Schaller}, \citenamefont {Brandes},\ and\
  \citenamefont {Esposito}}]{strasberg2013thermodynamics}%
  \BibitemOpen
  \bibfield  {author} {\bibinfo {author} {\bibfnamefont {P.}~\bibnamefont
  {Strasberg}}, \bibinfo {author} {\bibfnamefont {G.}~\bibnamefont {Schaller}},
  \bibinfo {author} {\bibfnamefont {T.}~\bibnamefont {Brandes}},\ and\ \bibinfo
  {author} {\bibfnamefont {M.}~\bibnamefont {Esposito}},\ }\bibfield  {title}
  {\bibinfo {title} {Thermodynamics of a physical model implementing a
  {M}axwell demon},\ }\href@noop {} {\bibfield  {journal} {\bibinfo  {journal}
  {Phys. Rev. Lett.}\ }\textbf {\bibinfo {volume} {110}},\ \bibinfo {pages}
  {040601} (\bibinfo {year} {2013})}\BibitemShut {NoStop}%
\bibitem [{\citenamefont {Horowitz}\ and\ \citenamefont
  {Esposito}(2014)}]{horowitz2014thermodynamics}%
  \BibitemOpen
  \bibfield  {author} {\bibinfo {author} {\bibfnamefont {J.~M.}\ \bibnamefont
  {Horowitz}}\ and\ \bibinfo {author} {\bibfnamefont {M.}~\bibnamefont
  {Esposito}},\ }\bibfield  {title} {\bibinfo {title} {Thermodynamics with
  continuous information flow},\ }\href@noop {} {\bibfield  {journal} {\bibinfo
   {journal} {Phys. Rev. X}\ }\textbf {\bibinfo {volume} {4}},\ \bibinfo
  {pages} {031015} (\bibinfo {year} {2014})}\BibitemShut {NoStop}%
\bibitem [{\citenamefont {Deffner}\ and\ \citenamefont
  {Jarzynski}(2013)}]{deffner2013information}%
  \BibitemOpen
  \bibfield  {author} {\bibinfo {author} {\bibfnamefont {S.}~\bibnamefont
  {Deffner}}\ and\ \bibinfo {author} {\bibfnamefont {C.}~\bibnamefont
  {Jarzynski}},\ }\bibfield  {title} {\bibinfo {title} {Information processing
  and the second law of thermodynamics: An inclusive, {H}amiltonian approach},\
  }\href@noop {} {\bibfield  {journal} {\bibinfo  {journal} {Phys. Rev. X}\
  }\textbf {\bibinfo {volume} {3}},\ \bibinfo {pages} {041003} (\bibinfo {year}
  {2013})}\BibitemShut {NoStop}%
\bibitem [{\citenamefont {Tasaki}(2013)}]{tasaki2013unified}%
  \BibitemOpen
  \bibfield  {author} {\bibinfo {author} {\bibfnamefont {H.}~\bibnamefont
  {Tasaki}},\ }\href@noop {} {\bibinfo {title} {Unified {J}arzynski and
  {S}agawa-{U}eda relations for {M}axwell's demon}} (\bibinfo {year} {2013}),\
  \Eprint {https://arxiv.org/abs/1308.3776 [cond-mat.stat-mech]}
  {arXiv:1308.3776 [cond-mat.stat-mech]} \BibitemShut {NoStop}%
\bibitem [{\citenamefont {Sagawa}\ and\ \citenamefont
  {Ueda}(2008)}]{sagawa2008second}%
  \BibitemOpen
  \bibfield  {author} {\bibinfo {author} {\bibfnamefont {T.}~\bibnamefont
  {Sagawa}}\ and\ \bibinfo {author} {\bibfnamefont {M.}~\bibnamefont {Ueda}},\
  }\bibfield  {title} {\bibinfo {title} {Second law of thermodynamics with
  discrete quantum feedback control},\ }\href@noop {} {\bibfield  {journal}
  {\bibinfo  {journal} {Phys. Rev. Lett.}\ }\textbf {\bibinfo {volume} {100}},\
  \bibinfo {pages} {080403} (\bibinfo {year} {2008})}\BibitemShut {NoStop}%
\bibitem [{\citenamefont {Hasegawa}\ \emph {et~al.}(2010)\citenamefont
  {Hasegawa}, \citenamefont {Ishikawa}, \citenamefont {Takara},\ and\
  \citenamefont {Driebe}}]{hasegawa2010generalization}%
  \BibitemOpen
  \bibfield  {author} {\bibinfo {author} {\bibfnamefont {H.-H.}\ \bibnamefont
  {Hasegawa}}, \bibinfo {author} {\bibfnamefont {J.}~\bibnamefont {Ishikawa}},
  \bibinfo {author} {\bibfnamefont {K.}~\bibnamefont {Takara}},\ and\ \bibinfo
  {author} {\bibfnamefont {D.}~\bibnamefont {Driebe}},\ }\bibfield  {title}
  {\bibinfo {title} {Generalization of the second law for a nonequilibrium
  initial state},\ }\href@noop {} {\bibfield  {journal} {\bibinfo  {journal}
  {Phys. Lett. A}\ }\textbf {\bibinfo {volume} {374}},\ \bibinfo {pages} {1001}
  (\bibinfo {year} {2010})}\BibitemShut {NoStop}%
\bibitem [{\citenamefont {Ozawa}(1988)}]{ozawa1988measurement}%
  \BibitemOpen
  \bibfield  {author} {\bibinfo {author} {\bibfnamefont {M.}~\bibnamefont
  {Ozawa}},\ }\bibfield  {title} {\bibinfo {title} {Measurement breaking the
  standard quantum limit for free-mass position},\ }\href@noop {} {\bibfield
  {journal} {\bibinfo  {journal} {Phys. Rev. Lett.}\ }\textbf {\bibinfo
  {volume} {60}},\ \bibinfo {pages} {385} (\bibinfo {year} {1988})}\BibitemShut
  {NoStop}%
\bibitem [{\citenamefont {Jacobs}\ and\ \citenamefont
  {Steck}(2006)}]{jacobs2006straightforward}%
  \BibitemOpen
  \bibfield  {author} {\bibinfo {author} {\bibfnamefont {K.}~\bibnamefont
  {Jacobs}}\ and\ \bibinfo {author} {\bibfnamefont {D.~A.}\ \bibnamefont
  {Steck}},\ }\bibfield  {title} {\bibinfo {title} {A straightforward
  introduction to continuous quantum measurement},\ }\href@noop {} {\bibfield
  {journal} {\bibinfo  {journal} {Contemp. Phys.}\ }\textbf {\bibinfo {volume}
  {47}},\ \bibinfo {pages} {279} (\bibinfo {year} {2006})}\BibitemShut
  {NoStop}%
\bibitem [{\citenamefont {Arnol'd}(1989)}]{arnol1989mathematical}%
  \BibitemOpen
  \bibfield  {author} {\bibinfo {author} {\bibfnamefont {V.~I.}\ \bibnamefont
  {Arnol'd}},\ }\href@noop {} {\emph {\bibinfo {title} {Mathematical methods of
  classical mechanics}}},\ \bibinfo {edition} {2nd}\ ed.\ (\bibinfo
  {publisher} {Springer},\ \bibinfo {year} {1989})\BibitemShut {NoStop}%
\bibitem [{\citenamefont {Libermann}\ and\ \citenamefont
  {Marle}(2012)}]{libermann2012symplectic}%
  \BibitemOpen
  \bibfield  {author} {\bibinfo {author} {\bibfnamefont {P.}~\bibnamefont
  {Libermann}}\ and\ \bibinfo {author} {\bibfnamefont {C.-M.}\ \bibnamefont
  {Marle}},\ }\href@noop {} {\emph {\bibinfo {title} {Symplectic geometry and
  analytical mechanics}}},\ Vol.~\bibinfo {volume} {35}\ (\bibinfo  {publisher}
  {Springer Science \& Business Media},\ \bibinfo {year} {2012})\BibitemShut
  {NoStop}%
\bibitem [{\citenamefont {Ozawa}(1993)}]{ozawa1993canonical}%
  \BibitemOpen
  \bibfield  {author} {\bibinfo {author} {\bibfnamefont {M.}~\bibnamefont
  {Ozawa}},\ }\bibfield  {title} {\bibinfo {title} {Canonical approximate
  quantum measurements},\ }\href@noop {} {\bibfield  {journal} {\bibinfo
  {journal} {J. Math. Phys.}\ }\textbf {\bibinfo {volume} {34}},\ \bibinfo
  {pages} {5596} (\bibinfo {year} {1993})}\BibitemShut {NoStop}%
\bibitem [{\citenamefont {Whittaker}(1959)}]{whittaker1959treatise}%
  \BibitemOpen
  \bibfield  {author} {\bibinfo {author} {\bibfnamefont {E.~T.}\ \bibnamefont
  {Whittaker}},\ }\href@noop {} {\emph {\bibinfo {title} {A treatise on the
  analytical dynamics of particles and rigid bodies}}},\ \bibinfo {edition}
  {4th}\ ed.\ (\bibinfo  {publisher} {Cambridge University Press, London},\
  \bibinfo {year} {1959})\ \bibinfo {note} {section 192}\BibitemShut {NoStop}%
\bibitem [{\citenamefont {Williamson}(1936)}]{williamson1936algebraic}%
  \BibitemOpen
  \bibfield  {author} {\bibinfo {author} {\bibfnamefont {J.}~\bibnamefont
  {Williamson}},\ }\bibfield  {title} {\bibinfo {title} {On the algebraic
  problem concerning the normal forms of linear dynamical systems},\
  }\href@noop {} {\bibfield  {journal} {\bibinfo  {journal} {Am. J. Math.}\
  }\textbf {\bibinfo {volume} {58}},\ \bibinfo {pages} {141} (\bibinfo {year}
  {1936})}\BibitemShut {NoStop}%
\bibitem [{\citenamefont {Kardar}(2007)}]{kardar2007statistical}%
  \BibitemOpen
  \bibfield  {author} {\bibinfo {author} {\bibfnamefont {M.}~\bibnamefont
  {Kardar}},\ }\href@noop {} {\emph {\bibinfo {title} {Statistical physics of
  particles}}}\ (\bibinfo  {publisher} {Cambridge University Press},\ \bibinfo
  {year} {2007})\BibitemShut {NoStop}%
\bibitem [{\citenamefont {Bismut}(1981)}]{bismut1981mecanique}%
  \BibitemOpen
  \bibfield  {author} {\bibinfo {author} {\bibfnamefont {J.-M.}\ \bibnamefont
  {Bismut}},\ }\href@noop {} {\emph {\bibinfo {title} {M{\'e}canique
  al{\'e}atoire}}},\ \bibinfo {series} {Lecture Notes in Mathematics}, Vol.\
  \bibinfo {volume} {866}\ (\bibinfo  {publisher} {Springer-Verlag, Berlin},\
  \bibinfo {year} {1981})\BibitemShut {NoStop}%
\bibitem [{\citenamefont {Parrondo}\ \emph {et~al.}(1990)\citenamefont
  {Parrondo}, \citenamefont {Ma{\~n}as},\ and\ \citenamefont {de~la
  Rubia}}]{parrondo1990geometrical}%
  \BibitemOpen
  \bibfield  {author} {\bibinfo {author} {\bibfnamefont {J.~M.~R.}\
  \bibnamefont {Parrondo}}, \bibinfo {author} {\bibfnamefont {M.}~\bibnamefont
  {Ma{\~n}as}},\ and\ \bibinfo {author} {\bibfnamefont {F.~J.}\ \bibnamefont
  {de~la Rubia}},\ }\bibfield  {title} {\bibinfo {title} {Geometrical treatment
  of systems driven by coloured noise},\ }\href@noop {} {\bibfield  {journal}
  {\bibinfo  {journal} {J. Phys. A-Math Gen}\ }\textbf {\bibinfo {volume}
  {23}},\ \bibinfo {pages} {2363} (\bibinfo {year} {1990})}\BibitemShut
  {NoStop}%
\bibitem [{\citenamefont {Ozawa}(2003)}]{ozawa2003universally}%
  \BibitemOpen
  \bibfield  {author} {\bibinfo {author} {\bibfnamefont {M.}~\bibnamefont
  {Ozawa}},\ }\bibfield  {title} {\bibinfo {title} {Universally valid
  reformulation of the {H}eisenberg uncertainty principle on noise and
  disturbance in measurement},\ }\href@noop {} {\bibfield  {journal} {\bibinfo
  {journal} {Phys. Rev. A}\ }\textbf {\bibinfo {volume} {67}},\ \bibinfo
  {pages} {042105} (\bibinfo {year} {2003})}\BibitemShut {NoStop}%
\bibitem [{\citenamefont {Erhart}\ \emph {et~al.}(2012)\citenamefont {Erhart},
  \citenamefont {Sponar}, \citenamefont {Sulyok}, \citenamefont {Badurek},
  \citenamefont {Ozawa},\ and\ \citenamefont
  {Hasegawa}}]{erhart2012experimental}%
  \BibitemOpen
  \bibfield  {author} {\bibinfo {author} {\bibfnamefont {J.}~\bibnamefont
  {Erhart}}, \bibinfo {author} {\bibfnamefont {S.}~\bibnamefont {Sponar}},
  \bibinfo {author} {\bibfnamefont {G.}~\bibnamefont {Sulyok}}, \bibinfo
  {author} {\bibfnamefont {G.}~\bibnamefont {Badurek}}, \bibinfo {author}
  {\bibfnamefont {M.}~\bibnamefont {Ozawa}},\ and\ \bibinfo {author}
  {\bibfnamefont {Y.}~\bibnamefont {Hasegawa}},\ }\bibfield  {title} {\bibinfo
  {title} {Experimental demonstration of a universally valid error--disturbance
  uncertainty relation in spin measurements},\ }\href@noop {} {\bibfield
  {journal} {\bibinfo  {journal} {Nat. Phys.}\ }\textbf {\bibinfo {volume}
  {8}},\ \bibinfo {pages} {185} (\bibinfo {year} {2012})}\BibitemShut {NoStop}%
\bibitem [{\citenamefont {Baek}\ \emph {et~al.}(2013)\citenamefont {Baek},
  \citenamefont {Kaneda}, \citenamefont {Ozawa},\ and\ \citenamefont
  {Edamatsu}}]{baek2013experimental}%
  \BibitemOpen
  \bibfield  {author} {\bibinfo {author} {\bibfnamefont {S.-Y.}\ \bibnamefont
  {Baek}}, \bibinfo {author} {\bibfnamefont {F.}~\bibnamefont {Kaneda}},
  \bibinfo {author} {\bibfnamefont {M.}~\bibnamefont {Ozawa}},\ and\ \bibinfo
  {author} {\bibfnamefont {K.}~\bibnamefont {Edamatsu}},\ }\bibfield  {title}
  {\bibinfo {title} {Experimental violation and reformulation of the
  {H}eisenberg's error-disturbance uncertainty relation},\ }\href@noop {}
  {\bibfield  {journal} {\bibinfo  {journal} {Sci. Rep.}\ }\textbf {\bibinfo
  {volume} {3}},\ \bibinfo {pages} {2221} (\bibinfo {year} {2013})}\BibitemShut
  {NoStop}%
\bibitem [{\citenamefont {Fujikawa}(2012)}]{fujikawa2012universally}%
  \BibitemOpen
  \bibfield  {author} {\bibinfo {author} {\bibfnamefont {K.}~\bibnamefont
  {Fujikawa}},\ }\bibfield  {title} {\bibinfo {title} {Universally valid
  {H}eisenberg uncertainty relation},\ }\href@noop {} {\bibfield  {journal}
  {\bibinfo  {journal} {Phys. Rev. A}\ }\textbf {\bibinfo {volume} {85}},\
  \bibinfo {pages} {062117} (\bibinfo {year} {2012})}\BibitemShut {NoStop}%
\bibitem [{\citenamefont {Busch}\ \emph {et~al.}(2013)\citenamefont {Busch},
  \citenamefont {Lahti},\ and\ \citenamefont {Werner}}]{busch2013proof}%
  \BibitemOpen
  \bibfield  {author} {\bibinfo {author} {\bibfnamefont {P.}~\bibnamefont
  {Busch}}, \bibinfo {author} {\bibfnamefont {P.}~\bibnamefont {Lahti}},\ and\
  \bibinfo {author} {\bibfnamefont {R.~F.}\ \bibnamefont {Werner}},\ }\bibfield
   {title} {\bibinfo {title} {Proof of {H}eisenberg’s error-disturbance
  relation},\ }\href@noop {} {\bibfield  {journal} {\bibinfo  {journal} {Phys.
  Rev. Lett.}\ }\textbf {\bibinfo {volume} {111}},\ \bibinfo {pages} {160405}
  (\bibinfo {year} {2013})}\BibitemShut {NoStop}%
\bibitem [{\citenamefont {Dressel}\ and\ \citenamefont
  {Nori}(2014)}]{dressel2014certainty}%
  \BibitemOpen
  \bibfield  {author} {\bibinfo {author} {\bibfnamefont {J.}~\bibnamefont
  {Dressel}}\ and\ \bibinfo {author} {\bibfnamefont {F.}~\bibnamefont {Nori}},\
  }\bibfield  {title} {\bibinfo {title} {Certainty in {H}eisenberg's
  uncertainty principle: Revisiting definitions for estimation errors and
  disturbance},\ }\href@noop {} {\bibfield  {journal} {\bibinfo  {journal}
  {Phys. Rev. A}\ }\textbf {\bibinfo {volume} {89}},\ \bibinfo {pages} {022106}
  (\bibinfo {year} {2014})}\BibitemShut {NoStop}%
\bibitem [{\citenamefont {Griffiths}(2005)}]{griffiths2005introduction}%
  \BibitemOpen
  \bibfield  {author} {\bibinfo {author} {\bibfnamefont {D.~J.}\ \bibnamefont
  {Griffiths}},\ }\href@noop {} {\emph {\bibinfo {title} {Introduction to
  quantum mechanics}}},\ \bibinfo {edition} {2nd}\ ed.\ (\bibinfo  {publisher}
  {Pearson Prentice Hall},\ \bibinfo {year} {2005})\BibitemShut {NoStop}%
\bibitem [{\citenamefont {Wigner}(1961)}]{wigner1961remarks}%
  \BibitemOpen
  \bibfield  {author} {\bibinfo {author} {\bibfnamefont {E.~P.}\ \bibnamefont
  {Wigner}},\ }\bibfield  {title} {\bibinfo {title} {Remarks on the mind body
  question},\ }in\ \href@noop {} {\emph {\bibinfo {booktitle} {The Scientist
  Speculates}}},\ \bibinfo {editor} {edited by\ \bibinfo {editor}
  {\bibfnamefont {I.~J.}\ \bibnamefont {Good}}}\ (\bibinfo  {publisher}
  {Heinemann, London},\ \bibinfo {year} {1961})\BibitemShut {NoStop}%
\bibitem [{\citenamefont {Bell}(1990)}]{bell1990against}%
  \BibitemOpen
  \bibfield  {author} {\bibinfo {author} {\bibfnamefont {J.}~\bibnamefont
  {Bell}},\ }\bibfield  {title} {\bibinfo {title} {Against ‘measurement’},\
  }\href@noop {} {\bibfield  {journal} {\bibinfo  {journal} {Phys. World}\
  }\textbf {\bibinfo {volume} {3}},\ \bibinfo {pages} {33} (\bibinfo {year}
  {1990})}\BibitemShut {NoStop}%
\bibitem [{\citenamefont {Nielsen}\ and\ \citenamefont
  {Chuang}(2002)}]{nielsen2002quantum}%
  \BibitemOpen
  \bibfield  {author} {\bibinfo {author} {\bibfnamefont {M.~A.}\ \bibnamefont
  {Nielsen}}\ and\ \bibinfo {author} {\bibfnamefont {I.}~\bibnamefont
  {Chuang}},\ }\href@noop {} {\emph {\bibinfo {title} {Quantum computation and
  quantum information}}}\ (\bibinfo  {publisher} {American Association of
  Physics Teachers},\ \bibinfo {year} {2002})\BibitemShut {NoStop}%
\bibitem [{\citenamefont {Busch}\ \emph {et~al.}(2014)\citenamefont {Busch},
  \citenamefont {Lahti},\ and\ \citenamefont {Werner}}]{busch2014colloquium}%
  \BibitemOpen
  \bibfield  {author} {\bibinfo {author} {\bibfnamefont {P.}~\bibnamefont
  {Busch}}, \bibinfo {author} {\bibfnamefont {P.}~\bibnamefont {Lahti}},\ and\
  \bibinfo {author} {\bibfnamefont {R.~F.}\ \bibnamefont {Werner}},\ }\bibfield
   {title} {\bibinfo {title} {Colloquium: Quantum root-mean-square error and
  measurement uncertainty relations},\ }\href@noop {} {\bibfield  {journal}
  {\bibinfo  {journal} {Rev. Mod. Phys.}\ }\textbf {\bibinfo {volume} {86}},\
  \bibinfo {pages} {1261} (\bibinfo {year} {2014})}\BibitemShut {NoStop}%
\bibitem [{\citenamefont {Hilgevoord}\ and\ \citenamefont
  {Uffink}(2024)}]{hilgevoord2024uncertainty}%
  \BibitemOpen
  \bibfield  {author} {\bibinfo {author} {\bibfnamefont {J.}~\bibnamefont
  {Hilgevoord}}\ and\ \bibinfo {author} {\bibfnamefont {J.}~\bibnamefont
  {Uffink}},\ }\bibfield  {title} {\bibinfo {title} {{The Uncertainty
  Principle}},\ }in\ \href@noop {} {\emph {\bibinfo {booktitle} {The {Stanford}
  Encyclopedia of Philosophy}}},\ \bibinfo {editor} {edited by\ \bibinfo
  {editor} {\bibfnamefont {E.~N.}\ \bibnamefont {Zalta}}\ and\ \bibinfo
  {editor} {\bibfnamefont {U.}~\bibnamefont {Nodelman}}}\ (\bibinfo
  {publisher} {Metaphysics Research Lab, Stanford University},\ \bibinfo {year}
  {2024})\ \bibinfo {edition} {{S}pring 2024}\ ed.\BibitemShut {Stop}%
\bibitem [{\citenamefont {Ozawa}(2019)}]{ozawa2019soundness}%
  \BibitemOpen
  \bibfield  {author} {\bibinfo {author} {\bibfnamefont {M.}~\bibnamefont
  {Ozawa}},\ }\bibfield  {title} {\bibinfo {title} {Soundness and completeness
  of quantum root-mean-square errors},\ }\href@noop {} {\bibfield  {journal}
  {\bibinfo  {journal} {Npj Quantum Inf.}\ }\textbf {\bibinfo {volume} {5}},\
  \bibinfo {pages} {1} (\bibinfo {year} {2019})}\BibitemShut {NoStop}%
\bibitem [{\citenamefont {Ozawa}(2002)}]{ozawa2002position}%
  \BibitemOpen
  \bibfield  {author} {\bibinfo {author} {\bibfnamefont {M.}~\bibnamefont
  {Ozawa}},\ }\bibfield  {title} {\bibinfo {title} {Position measuring
  interactions and the {H}eisenberg uncertainty principle},\ }\href@noop {}
  {\bibfield  {journal} {\bibinfo  {journal} {Phys. Lett. A}\ }\textbf
  {\bibinfo {volume} {299}},\ \bibinfo {pages} {1} (\bibinfo {year}
  {2002})}\BibitemShut {NoStop}%
\bibitem [{\citenamefont {Davies}\ and\ \citenamefont
  {Lewis}(1970)}]{davies1970operational}%
  \BibitemOpen
  \bibfield  {author} {\bibinfo {author} {\bibfnamefont {E.~B.}\ \bibnamefont
  {Davies}}\ and\ \bibinfo {author} {\bibfnamefont {J.~T.}\ \bibnamefont
  {Lewis}},\ }\bibfield  {title} {\bibinfo {title} {An operational approach to
  quantum probability},\ }\href@noop {} {\bibfield  {journal} {\bibinfo
  {journal} {Commun. Math. Phys.}\ }\textbf {\bibinfo {volume} {17}},\ \bibinfo
  {pages} {239} (\bibinfo {year} {1970})}\BibitemShut {NoStop}%
\bibitem [{\citenamefont {Kraus}(1971)}]{kraus1971general}%
  \BibitemOpen
  \bibfield  {author} {\bibinfo {author} {\bibfnamefont {K.}~\bibnamefont
  {Kraus}},\ }\bibfield  {title} {\bibinfo {title} {General state changes in
  quantum theory},\ }\href@noop {} {\bibfield  {journal} {\bibinfo  {journal}
  {Ann. Phys.}\ }\textbf {\bibinfo {volume} {64}},\ \bibinfo {pages} {311}
  (\bibinfo {year} {1971})}\BibitemShut {NoStop}%
\bibitem [{\citenamefont {Ozawa}(1984)}]{ozawa1984quantum}%
  \BibitemOpen
  \bibfield  {author} {\bibinfo {author} {\bibfnamefont {M.}~\bibnamefont
  {Ozawa}},\ }\bibfield  {title} {\bibinfo {title} {Quantum measuring processes
  of continuous observables},\ }\href@noop {} {\bibfield  {journal} {\bibinfo
  {journal} {J. Math. Phys.}\ }\textbf {\bibinfo {volume} {25}},\ \bibinfo
  {pages} {79} (\bibinfo {year} {1984})}\BibitemShut {NoStop}%
\bibitem [{\citenamefont {Ozawa}(2004)}]{ozawa2004uncertainty}%
  \BibitemOpen
  \bibfield  {author} {\bibinfo {author} {\bibfnamefont {M.}~\bibnamefont
  {Ozawa}},\ }\bibfield  {title} {\bibinfo {title} {Uncertainty relations for
  noise and disturbance in generalized quantum measurements},\ }\href@noop {}
  {\bibfield  {journal} {\bibinfo  {journal} {Ann. Phys.-New York}\ }\textbf
  {\bibinfo {volume} {311}},\ \bibinfo {pages} {350} (\bibinfo {year}
  {2004})}\BibitemShut {NoStop}%
\bibitem [{\citenamefont {Belavkin}(1992)}]{belavkin1992quantum}%
  \BibitemOpen
  \bibfield  {author} {\bibinfo {author} {\bibfnamefont {V.~P.}\ \bibnamefont
  {Belavkin}},\ }\bibfield  {title} {\bibinfo {title} {Quantum continual
  measurements and a posteriori collapse on {CCR}},\ }\href@noop {} {\bibfield
  {journal} {\bibinfo  {journal} {Commun. Math. Phys.}\ }\textbf {\bibinfo
  {volume} {146}},\ \bibinfo {pages} {611} (\bibinfo {year}
  {1992})}\BibitemShut {NoStop}%
\bibitem [{\citenamefont {Ozawa}(1990)}]{ozawa1990quantum}%
  \BibitemOpen
  \bibfield  {author} {\bibinfo {author} {\bibfnamefont {M.}~\bibnamefont
  {Ozawa}},\ }\bibfield  {title} {\bibinfo {title} {Quantum-mechanical models
  of position measurements},\ }\href@noop {} {\bibfield  {journal} {\bibinfo
  {journal} {Phys. Rev. A}\ }\textbf {\bibinfo {volume} {41}},\ \bibinfo
  {pages} {1735} (\bibinfo {year} {1990})}\BibitemShut {NoStop}%
\bibitem [{\citenamefont {Gromov}(1985)}]{gromov1985pseudo}%
  \BibitemOpen
  \bibfield  {author} {\bibinfo {author} {\bibfnamefont {M.}~\bibnamefont
  {Gromov}},\ }\bibfield  {title} {\bibinfo {title} {Pseudo holomorphic curves
  in symplectic manifolds},\ }\href@noop {} {\bibfield  {journal} {\bibinfo
  {journal} {Invent. Math.}\ }\textbf {\bibinfo {volume} {82}},\ \bibinfo
  {pages} {307} (\bibinfo {year} {1985})}\BibitemShut {NoStop}%
\bibitem [{\citenamefont {Hofer}\ and\ \citenamefont
  {Zehnder}(1994)}]{hofer1994symplectic}%
  \BibitemOpen
  \bibfield  {author} {\bibinfo {author} {\bibfnamefont {H.}~\bibnamefont
  {Hofer}}\ and\ \bibinfo {author} {\bibfnamefont {E.}~\bibnamefont
  {Zehnder}},\ }\href@noop {} {\emph {\bibinfo {title} {Symplectic invariants
  and {H}amiltonian dynamics}}}\ (\bibinfo  {publisher} {Birkh{\"a}user},\
  \bibinfo {year} {1994})\BibitemShut {NoStop}%
\bibitem [{\citenamefont {Hsiao}\ and\ \citenamefont
  {Scheeres}(2006)}]{hsiao2006fundamental}%
  \BibitemOpen
  \bibfield  {author} {\bibinfo {author} {\bibfnamefont {F.-Y.}\ \bibnamefont
  {Hsiao}}\ and\ \bibinfo {author} {\bibfnamefont {D.~J.}\ \bibnamefont
  {Scheeres}},\ }\bibfield  {title} {\bibinfo {title} {Fundamental constraints
  on uncertainty evolution in {H}amiltonian systems},\ }in\ \href@noop {}
  {\emph {\bibinfo {booktitle} {P. Amer. Contr. Conf.}}}\ (\bibinfo
  {organization} {IEEE},\ \bibinfo {year} {2006})\ pp.\ \bibinfo {pages}
  {5033--5038}\BibitemShut {NoStop}%
\bibitem [{\citenamefont {de~Gosson}(2009)}]{de2009symplectic}%
  \BibitemOpen
  \bibfield  {author} {\bibinfo {author} {\bibfnamefont {M.~A.}\ \bibnamefont
  {de~Gosson}},\ }\bibfield  {title} {\bibinfo {title} {The symplectic camel
  and the uncertainty principle: The tip of an iceberg?},\ }\href@noop {}
  {\bibfield  {journal} {\bibinfo  {journal} {Found. Phys.}\ }\textbf {\bibinfo
  {volume} {39}},\ \bibinfo {pages} {194} (\bibinfo {year} {2009})}\BibitemShut
  {NoStop}%
\bibitem [{\citenamefont {Wiseman}(1996)}]{wiseman1996quantum}%
  \BibitemOpen
  \bibfield  {author} {\bibinfo {author} {\bibfnamefont {H.}~\bibnamefont
  {Wiseman}},\ }\bibfield  {title} {\bibinfo {title} {Quantum trajectories and
  quantum measurement theory},\ }\href@noop {} {\bibfield  {journal} {\bibinfo
  {journal} {Quantum Semicl. Opt.}\ }\textbf {\bibinfo {volume} {8}},\ \bibinfo
  {pages} {205} (\bibinfo {year} {1996})}\BibitemShut {NoStop}%
\bibitem [{\citenamefont {Einstein}\ \emph {et~al.}(1935)\citenamefont
  {Einstein}, \citenamefont {Podolsky},\ and\ \citenamefont
  {Rosen}}]{einstein1935can}%
  \BibitemOpen
  \bibfield  {author} {\bibinfo {author} {\bibfnamefont {A.}~\bibnamefont
  {Einstein}}, \bibinfo {author} {\bibfnamefont {B.}~\bibnamefont {Podolsky}},\
  and\ \bibinfo {author} {\bibfnamefont {N.}~\bibnamefont {Rosen}},\ }\bibfield
   {title} {\bibinfo {title} {Can quantum-mechanical description of physical
  reality be considered complete?},\ }\href@noop {} {\bibfield  {journal}
  {\bibinfo  {journal} {Phys. Rev.}\ }\textbf {\bibinfo {volume} {47}},\
  \bibinfo {pages} {777} (\bibinfo {year} {1935})}\BibitemShut {NoStop}%
\bibitem [{\citenamefont {Harrigan}\ and\ \citenamefont
  {Spekkens}(2010)}]{harrigan2010einstein}%
  \BibitemOpen
  \bibfield  {author} {\bibinfo {author} {\bibfnamefont {N.}~\bibnamefont
  {Harrigan}}\ and\ \bibinfo {author} {\bibfnamefont {R.~W.}\ \bibnamefont
  {Spekkens}},\ }\bibfield  {title} {\bibinfo {title} {Einstein,
  incompleteness, and the epistemic view of quantum states},\ }\href@noop {}
  {\bibfield  {journal} {\bibinfo  {journal} {Found. Phys.}\ }\textbf {\bibinfo
  {volume} {40}},\ \bibinfo {pages} {125} (\bibinfo {year} {2010})}\BibitemShut
  {NoStop}%
\bibitem [{\citenamefont {Fuchs}(2002)}]{fuchs2002quantum}%
  \BibitemOpen
  \bibfield  {author} {\bibinfo {author} {\bibfnamefont {C.~A.}\ \bibnamefont
  {Fuchs}},\ }\href@noop {} {\bibinfo {title} {Quantum mechanics as quantum
  information (and only a little more)}} (\bibinfo {year} {2002}),\ \Eprint
  {https://arxiv.org/abs/quant-ph/0205039} {arXiv:quant-ph/0205039}
  \BibitemShut {NoStop}%
\bibitem [{\citenamefont {Caves}\ \emph {et~al.}(2002)\citenamefont {Caves},
  \citenamefont {Fuchs},\ and\ \citenamefont {Schack}}]{caves2002unknown}%
  \BibitemOpen
  \bibfield  {author} {\bibinfo {author} {\bibfnamefont {C.~M.}\ \bibnamefont
  {Caves}}, \bibinfo {author} {\bibfnamefont {C.~A.}\ \bibnamefont {Fuchs}},\
  and\ \bibinfo {author} {\bibfnamefont {R.}~\bibnamefont {Schack}},\
  }\bibfield  {title} {\bibinfo {title} {Unknown quantum states: The quantum
  {de Finetti} representation},\ }\href@noop {} {\bibfield  {journal} {\bibinfo
   {journal} {J. Math. Phys.}\ }\textbf {\bibinfo {volume} {43}},\ \bibinfo
  {pages} {4537} (\bibinfo {year} {2002})}\BibitemShut {NoStop}%
\bibitem [{\citenamefont {Spekkens}(2007)}]{spekkens2007evidence}%
  \BibitemOpen
  \bibfield  {author} {\bibinfo {author} {\bibfnamefont {R.~W.}\ \bibnamefont
  {Spekkens}},\ }\bibfield  {title} {\bibinfo {title} {Evidence for the
  epistemic view of quantum states: A toy theory},\ }\href@noop {} {\bibfield
  {journal} {\bibinfo  {journal} {Phys. Rev. A}\ }\textbf {\bibinfo {volume}
  {75}},\ \bibinfo {pages} {032110} (\bibinfo {year} {2007})}\BibitemShut
  {NoStop}%
\bibitem [{\citenamefont {Spekkens}(2016)}]{spekkens2016quasi}%
  \BibitemOpen
  \bibfield  {author} {\bibinfo {author} {\bibfnamefont {R.~W.}\ \bibnamefont
  {Spekkens}},\ }\bibfield  {title} {\bibinfo {title} {Quasi-quantization:
  Classical statistical theories with an epistemic restriction},\ }in\
  \href@noop {} {\emph {\bibinfo {booktitle} {Quantum Theory: Informational
  Foundations and Foils}}}\ (\bibinfo  {publisher} {Springer},\ \bibinfo {year}
  {2016})\ pp.\ \bibinfo {pages} {83--135}\BibitemShut {NoStop}%
\bibitem [{\citenamefont {Bartlett}\ \emph {et~al.}(2012)\citenamefont
  {Bartlett}, \citenamefont {Rudolph},\ and\ \citenamefont
  {Spekkens}}]{bartlett2012reconstruction}%
  \BibitemOpen
  \bibfield  {author} {\bibinfo {author} {\bibfnamefont {S.~D.}\ \bibnamefont
  {Bartlett}}, \bibinfo {author} {\bibfnamefont {T.}~\bibnamefont {Rudolph}},\
  and\ \bibinfo {author} {\bibfnamefont {R.~W.}\ \bibnamefont {Spekkens}},\
  }\bibfield  {title} {\bibinfo {title} {Reconstruction of {G}aussian quantum
  mechanics from {L}iouville mechanics with an epistemic restriction},\
  }\href@noop {} {\bibfield  {journal} {\bibinfo  {journal} {Phys. Rev. A}\
  }\textbf {\bibinfo {volume} {86}},\ \bibinfo {pages} {012103} (\bibinfo
  {year} {2012})}\BibitemShut {NoStop}%
\bibitem [{\citenamefont {Frank}(2017)}]{frank2017foundations}%
  \BibitemOpen
  \bibfield  {author} {\bibinfo {author} {\bibfnamefont {M.~P.}\ \bibnamefont
  {Frank}},\ }\bibfield  {title} {\bibinfo {title} {Foundations of generalized
  reversible computing},\ }in\ \href@noop {} {\emph {\bibinfo {booktitle}
  {International Conference on Reversible Computation}}}\ (\bibinfo
  {organization} {Springer},\ \bibinfo {year} {2017})\ pp.\ \bibinfo {pages}
  {19--34}\BibitemShut {NoStop}%
\bibitem [{\citenamefont {Arnol'd}\ and\ \citenamefont
  {Givental'}(1990)}]{arnold1990symplectic}%
  \BibitemOpen
  \bibfield  {author} {\bibinfo {author} {\bibfnamefont {V.~I.}\ \bibnamefont
  {Arnol'd}}\ and\ \bibinfo {author} {\bibfnamefont {A.~B.}\ \bibnamefont
  {Givental'}},\ }\bibfield  {title} {\bibinfo {title} {Symplectic geometry},\
  }in\ \href@noop {} {\emph {\bibinfo {booktitle} {Dynamical Systems IV:
  Symplectic geometry and its applications}}}\ (\bibinfo  {publisher} {Berlin:
  Springer-Verlag},\ \bibinfo {year} {1990})\ pp.\ \bibinfo {pages}
  {1--136}\BibitemShut {NoStop}%
\bibitem [{\citenamefont {Kirk}(1970)}]{kirk1970optimal}%
  \BibitemOpen
  \bibfield  {author} {\bibinfo {author} {\bibfnamefont {D.~E.}\ \bibnamefont
  {Kirk}},\ }\href@noop {} {\emph {\bibinfo {title} {Optimal control theory: An
  introduction}}}\ (\bibinfo  {publisher} {Prentice-Hall, Inc.},\ \bibinfo
  {year} {1970})\BibitemShut {NoStop}%
\end{thebibliography}%

\end{document}